\documentclass[10pt, conference, letterpaper]{IEEEtran}

\usepackage[T1]{fontenc}
\usepackage[utf8]{inputenc}
\usepackage{xspace}
\usepackage[textsize=tiny]{todonotes}
\usepackage[binary-units]{siunitx}
\usepackage[USenglish]{babel}
\usepackage{amsmath}
\usepackage{amssymb}
\makeatletter
\let\MYcaption\@makecaption
\makeatother
\usepackage[font=footnotesize]{caption}
\usepackage[font=footnotesize]{subcaption}
\usepackage{enumitem}
\usepackage{float}

\usetikzlibrary{arrows,shapes,positioning,calc}

\makeatletter
\let\MYcaption\@makecaption
\makeatother

\usepackage[nolist]{acronym}

\usepackage[capitalise,nameinlink]{cleveref}
\crefname{section}{Sec.}{Sec.}
\Crefname{section}{Section}{Sections}
\crefname{equation}{eq.}{eq.}
\crefname{figure}{Fig.}{Fig.s}
\Crefname{figure}{Figure}{Figures}

\usepackage[%
 backend=biber,
 style=ieee,
 url=false,
 isbn=false,
 doi=false,
 maxcitenames=1,
 mincitenames=1,
 maxbibnames=5,
]{biblatex}
\renewcommand*{\bibfont}{\footnotesize}

\AtEveryBibitem{\clearfield{note}}
\AtEveryBibitem{\clearlist{location}}
\AtEveryBibitem{\clearfield{pages}}
\AtEveryBibitem{\clearfield{series}}
\AtEveryBibitem{\clearfield{month}}

\newcommand*{\eg}{e.\,g.\@\xspace}
\newcommand*{\ie}{i.\,e.\@\xspace}
\newcommand{\R}{\mathbb{R}}
\newcommand{\N}{\mathbb{N}}
\newcommand{\Pb}{\mathbb{P}}
\newcommand{\Ex}{\mathbb{E}}

\DeclareMathOperator{\addr}{addr}
\DeclareMathOperator{\dist}{dist}
\newcommand{\node}{v}
\newcommand{\dataitem}{d}
\newcommand{\providers}{P(\dataitem)}

\begin{document}
\title{Mapping the \acl{IPFS}}

\begin{acronym}[Derp]
\acro{DHT}[DHT]{distributed hash table}
\acro{IPFS}[IPFS]{Interplanetary Filesystem}
\acro{SFS}[SFS]{Self-Certifying Filesystem}
\acro{IPNS}[IPNS]{Interplanetary Namesystem}
\end{acronym}

\author{
\IEEEauthorblockN{Sebastian Henningsen \qquad Martin Florian \qquad Sebastian Rust \qquad Björn Scheuermann}
\IEEEauthorblockA{~\\
Weizenbaum-Institute for the Networked Society\\
/ Humboldt-Universität zu Berlin\\Berlin, Germany\\
}}
\maketitle
\begin{abstract}
The \ac{IPFS} is a distributed data storage service frequently used by blockchain applications and for sharing content in a censorship-resistant manner.
Data is distributed within an open set of peers using a Kademlia-based \ac{DHT}.
In this paper, we study the structure of the resulting overlay network, as it significantly influences the robustness and performance of \ac{IPFS}.
We monitor and systematically crawl \ac{IPFS}' \ac{DHT} towards mapping the \ac{IPFS} overlay network.
Our measurements found an average of 44474 nodes at every given time.
At least \SI[detect-weight=true, detect-family=true]{52.19}{\percent} of these reside behind a NAT and are not reachable from the outside, suggesting that a large share of the network is operated by private individuals on an as-needed basis.
Based on our measurements and our analysis of the \ac{IPFS} code, we conclude that the topology of the \ac{IPFS} network is, in its current state, closer to an unstructured overlay network than it is to a classical \ac{DHT}.
While such a structure has benefits for robustness and the resistance against Sybil attacks,
it leaves room for improvement in terms of performance and query privacy.
\end{abstract}

\section{Introduction}
\label{sec:intro}

The \acl{IPFS}~\cite{DBLP:journals/corr/Benet14} is a community-developed peer-to-peer protocol and network providing public data storage services.
\ac{IPFS} is often cited as a fitting data storage solution for blockchain-based applications~\cite{li2018crowdbc, muhle2018survey, liu2017blockchain, tenorio2019towards, ascigil2019towards} and was previously used for mirroring censorship-threatened websites such as Wikipedia\footnote{https://github.com/ipfs/distributed-wikipedia-mirror}.
\ac{IPFS}' design is reminiscent to classical peer-to-peer systems such as filesharing networks~\cite{DBLP:conf/iptps/PouwelseGES05}.
Any Internet-enabled device can participate as an \ac{IPFS} node and nodes are operated without explicit economic incentives.
Unlike blockchain-based systems such as Bitcoin~\cite{nakamoto2008bitcoin}, data items (files, folders, ...) are not replicated globally.
Instead, each data item is stored by a small set of nodes, that item's \emph{providers}, who make the data item available to other peers.
Data items are addressed through immutable, cryptographically-generated names which are resolved to their providers through a \acl{DHT} based on Kademlia~\cite{DBLP:conf/iptps/MaymounkovM02}.

Given \ac{IPFS}' reported attractiveness as a building block for decentralized applications and
censorship circumvention, the question arises whether the network is actually suited to fulfill this role in terms of robustness and "decentrality".
We are particularity interested in the following questions:
\begin{itemize}
  \item What is possible with regards to mapping and monitoring the \ac{IPFS} overlay network (assessing its "health")?
  \item How many and what types of nodes participate in the \ac{IPFS} network? What kind of churn do they exbibit?
  \item How "decentralized" is the network -- in terms of overlay structure and the geographical distribution of nodes?
\end{itemize}

In this paper, we present the results of a comprehensive empirical study on the \ac{IPFS} overlay network.
As central components of our methodology, we collected connection data from several self-controlled monitoring nodes and repeatedly crawled the \ac{IPFS} \ac{DHT}.
We find that,
similar to other Kademlia-based systems~\cite{DBLP:conf/iscc/SalahRS14,DBLP:conf/iptps/PouwelseGES05},
connections corresponding to \ac{DHT} routing table entries
can be learned through carefully crafted, iterative peer discovery queries.
However, through measurements and studies of the \ac{IPFS} code base, we also uncover a number of differences to both previous systems and the design outlined in the \ac{IPFS} whitepaper~\cite{DBLP:journals/corr/Benet14}.
For example, in contrast to other Kademlia implementations, \ac{IPFS} establishes a connection with every peer it encounters and maintains a large number of connections that do not correspond to any \ac{DHT} routing table entries.
As a further surprise, and despite whitepaper claims to implementing mechanisms from \cite{DBLP:conf/icpads/BaumgartM07}, we find that no noteworthy protection against Sybil attacks is currently implemented in \ac{IPFS}.
The effect on security, however, is limited.
Thanks to \ac{IPFS}' unusual hoarding of overlay connections and the fact that requests are promiscuously sent to all direct peers,
content retrieval is possible even if an attacker can place Sybil nodes at key locations in the overlay.

Our contributions are threefold:
\begin{itemize}
  \item We give an overview on the \ac{IPFS} system based on white papers, public online discussions, and code.
    Notably, we describe the actual state of current implementations, and contrast it with the design documents.
  \item We run monitoring nodes with varying connectivity (fully reachable, behind NAT), some of which modified to accept an unlimited number of overlay connections.
    Among other things, this allows us to map the quantitative relationship between overlay connections and \ac{DHT} routing table entries, as influenced by node connectivity.
  \item We repeatedly crawled the \ac{IPFS} \ac{DHT}
    to obtain its topology,
    thereby also enumerating all \ac{DHT}-enabled nodes and their addresses.
\end{itemize}

\section{Related Work}
\label{sec:rw}

Peer-to-Peer networks have been studied extensively in the past, yielding various results on network crawling and
characterization~\cite{DBLP:journals/ton/SteinerEB09,DBLP:conf/iptps/MemonRGS09,4784962,DBLP:conf/infocom/StutzbachR06a,DBLP:conf/imc/SteinerEB07,6679883,6005906,DBLP:conf/iptps/SteinerBE07,DBLP:conf/pam/StutzbachR05}
with applications to real-world peer-to-peer systems like BitTorrent~\cite{DBLP:conf/iptps/PouwelseGES05} and KAD~\cite{DBLP:conf/imc/SteinerEB07}.
We extend this line of research by developing and performing a measurement study on \ac{IPFS} -- a highly popular data storage network with various reported applications (see, e.g., \cite{li2018crowdbc, liu2017blockchain, tenorio2019towards} for a sample of academic projects).

In their seminal work, \textcite{DBLP:conf/pam/StutzbachR05,DBLP:conf/infocom/StutzbachR06a} study requirements and pitfalls with regards to obtaining accurate snapshots of peer-to-peer overlays.
Specifically, they find that the duration of crawls should be as small as possible, to avoid distortions in the results due to churn.
\textcite{DBLP:conf/imc/SteinerEB07,DBLP:conf/iptps/SteinerBE07} crawled the KAD network to obtain the number of peers and their geographical distribution as well as inter-session times.
\textcite{DBLP:conf/iscc/SalahRS14} studied the graph-theoretical properties of KAD and contrasted their results with analytical considerations.
Similarly to KAD and other networks, the \ac{DHT} component of \ac{IPFS} is also, in principle, based on Kademlia~\cite{DBLP:conf/iptps/MaymounkovM02}.
We therefore build upon previously proposed crawling methods.
However, we also find that \ac{IPFS} differs substantially from more canonical Kademlia implementations, necessitating enhancements to existing measurement approaches.
We furthermore explore links in the \ac{IPFS} overlay that are not used for \ac{DHT} functionality.

A simple crawler for the \ac{IPFS} \ac{DHT} has been made available before\footnote{\url{https://github.com/vyzo/ipfs-crawl}} that aims at enumerating all nodes in the network.
For this paper, we developed a new crawler from scratch to capture the entire overlay topology.
It is optimized for short running times to ensure the correctness of snapshots\footnote{The code of our crawler can be found at \url{https://github.com/scriptkitty/ipfs-crawler}}.

The I/O performance of retrieving and storing data on \ac{IPFS} was studied in~\cite{DBLP:conf/iwqos/ShenLZW19,ascigil2019towards,7830696,confais2017object}, with interesting and partially contrasting results.
\textcite{ascigil2019towards} report high latencies, low throughput and high redundancy when retrieving data through \ac{IPFS}.
Similarly, \textcite{DBLP:conf/iwqos/ShenLZW19} report high latencies for large files and large variances in the transmission speed.
In~\cite{7830696,confais2017object}, the authors optimize and evaluate the performance of \ac{IPFS} in edge computing settings.
They report small latencies and high throughput when using the global \ac{DHT} as little as possible and running \ac{IPFS} in a private local network.
In contrast to these prior works, we focus on grasping the overlay structure and node composition of the public \ac{IPFS} network.
Furthermore, we give a comprehensive, code review-supported overview of \ac{IPFS}' ``network layer'',
revealing information not previously available in literature or documentation.

\section{The \acl{IPFS}}
\label{sec:ipfs-nutshell}

In the following, we describe key aspects of \ac{IPFS}' design and discuss particularities relevant to conducting a measurement study and interpreting its results.
It is worth noting that the development of \ac{IPFS} is ongoing, so that details of the design may change over time.
Here, we focus on inherent conceptual properties that change rarely in deployed protocols.

\subsection{In a Nutshell}
\label{subsec:tl;dr}

As a broad overview, the design of \ac{IPFS} can be summarized in the following way:

\begin{itemize}
  \item Data items are stored and served by data providers.
  \item References to data providers are stored in a \ac{DHT}.
  \item In \emph{addition} to \ac{DHT}-based provider discovery, data items are requested from \emph{all} connected overlay neighbors.
  \item \ac{DHT}: Kademlia over TCP (and other reliable transports), $k = 20$, no eviction mechanism and replacement buffers.
  \item In contrast to information in the \ac{IPFS} white paper~\cite{DBLP:journals/corr/Benet14}, no proposals of S/Kademlia~\cite{DBLP:conf/icpads/BaumgartM07} are implemented.
  \item Overlay connections can correspond to \ac{DHT} routing table (bucket) entries, but do not have to.
  \item By crawling the DHT we obtain a subset of all connections; we estimate the size of that subset in \cref{subsec:quality_of_crawl}.
\end{itemize}

\subsection{Node Identities and S/Kademlia}
\label{subsec:identities_skademlia}

Anyone can join the \ac{IPFS} overlay network, i.\,e., it is an open (permissionless) system with weak identities.
Nodes are identified by the hash of their public key, $H(k_{pub})$.
To ensure flexibility, \ac{IPFS} uses so-called ``multi-hashes'': a container format capable of supporting different hash functions.
A multi-hash adds meta information about the hash function and the digest length to a hash.
By default, \ac{IPFS} uses RSA2048 key pairs and SHA256 hashes.

Creating a new \ac{IPFS} identity is as simple as generating a new RSA key pair -- making the network highly susceptible to Sybil attacks~\cite{douceur2002sybil}.
Towards increasing the cost of Sybil attacks, the \ac{IPFS} white paper
\cite{DBLP:journals/corr/Benet14} suggests that the Proof-of-Work-based ID generation approach of S/Kademlia~\cite{DBLP:conf/icpads/BaumgartM07} is in use.
However, based on our careful review of the IPFS codebase, this is \emph{not} the case at the moment.
\ac{IPFS} currently implements no restriction on the generation of node IDs, neither are \ac{DHT} lookups carried out through multiple disjoint paths, as proposed in S/Kademlia.
IP address-based Sybil protection measures, such as limiting the number of connections to nodes from the same /24 subnet, are currently also \emph{not} in use.

On the network layer, \ac{IPFS} uses a concept of so-called ``multi-addresses''.
Similar to multi-hashes, these multi-addresses are capable of encoding a multitude of network and transport layer protocols.
Through these multi-addresses, a peer announces its connection capabilities, \eg, IPv4, IPv6, etc., and the addresses it can be reached from to the network.

\subsection{Data Storage: \acl{SFS}}
\label{subsec:sfs}

\ac{IPFS} uses a form of \ac{SFS} (originally introduced in~\cite{DBLP:phd/ndltd/Mazieres00}) to ensure the integrity of data throughout its delivery.
In \ac{IPFS}, each data item $\dataitem$ is assigned a unique immutable address that is the hash of its content, i.\,e. $\addr(\dataitem) = H(\dataitem)$.
Technically, folders and files are organized in a Merkle tree-like structure.
For example, a folder entry contains the hashes of all files in the folder.
In the end, the receiver can recognize whether received data was tampered with by comparing its
hash with the requested address.

Due to these hash structures, the address of data changes whenever the data itself is changed, which is not ideal in most circumstances.
As a partial remedy, \ac{IPFS} introduces the so-called \ac{IPNS}; a way of issuing self-certifying mutable addresses to data.
An address in \ac{IPNS} is the hash of a public key, containing a record that links to the hash of the desired data.
The hash is signed with the corresponding private key, making it easy to check the integrity and authenticity of the \ac{IPNS} record (hence, \emph{self-certifying}).
\ac{IPNS} addresses can only be updated by the issuer by publishing the update on the overlay through the \ac{DHT}.

\subsection{Content retrieval}
\label{subsec:content_retrieval}

Data items are usually stored at multiple nodes in the network.
Nodes store content because they are its original author or because they have recently retrieved it themselves.
Nodes normally \emph{serve} the data items they store, upon request.
The nodes that store a given data item are consequently referred to as that data item's \emph{providers}.

When an \ac{IPFS} node $\node$ wishes to retrieve a given data item $\dataitem$ (e.\,g., based on a user request), it follows two strategies \emph{in parallel}:
\begin{enumerate}
  \item Look up providers $\providers$ for $\dataitem$ in the \ac{DHT} (cf. \cref{subsec:ipfs_dht}), then request $\dataitem$ from members of $\providers$.
  \item Ask \emph{all} nodes it is currently connected to for $\dataitem$, using the \emph{Bitswap} subprotocol (cf. \cref{subsec:bitswap_exchange})
\end{enumerate}

\subsection{Content Retrieval: Kademlia \ac{DHT}}
\label{subsec:ipfs_dht}

\begin{figure}[htpb]
  \centering

\begin{tikzpicture}[inner sep=0.05cm, node distance = 0.5cm]
	\node[black, circle, draw] (v0) at (0, 3) {$v_0$};
	\node[below of=v0] (v1data) {$\{d_0, d_3\}$};

	\node[black, circle, draw] (v1) at (6, 3) {$v_1$};
	\node[below of=v1] (v2data) {$\{d_1\}$};

	\node[black, circle, draw] (v2) at (0, 0) {$v_2$};
	\node[below of=v2] (v3data) {$\{d_0, d_2, d_3\}$};

	\node[black, circle, draw] (v3) at (6, 0) {$v_3$};
	\node[below of=v3] (v4data) {$\{d_0\}$};

	\node[cloud, cloud puffs=7.3, cloud ignores aspect, align=left, draw] (cloudDHT) at (3, 1.25) {
                \\
		$d_0 \rightarrow \{v_0, v_2, v_3\}$\\
		$d_1 \rightarrow \{v_1\}$\\
		$d_2 \rightarrow \{v_2\}$\\
		$d_3 \rightarrow \{v_0, v_2\}$
	};
	\node[above=-0.9cm of cloudDHT] (dhtheading) {\textbf{DHT}};
\end{tikzpicture}
   \caption{Data items and the DHT.}
  \label{fig:dht-data}
\end{figure}
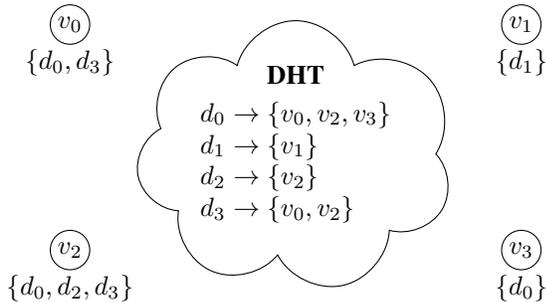

\ac{IPFS} uses a Kademlia-based \ac{DHT}.
Kademlia offers mechanisms for efficiently locating data records in a peer-to-peer network~\cite{DBLP:conf/iptps/MaymounkovM02}.
Each node in the \ac{IPFS} overlay network is uniquely identified by its node ID---the SHA-2 hash of its public key.
Data records are indexed and located using \emph{keys}.
Records in the \ac{IPFS} DHT are lists of data providers for data items stored in \ac{IPFS}, and the keys to these records are consequently the addresses of data items ($H(\dataitem)$ for a data item $\dataitem$).
An example scenario illustrating this relationship is presented in \cref{fig:dht-data}.
Nodes can be part of the \ac{IPFS} overlay network but choose to not participate in the \ac{DHT}, by acting as so-called \emph{client nodes}.
Therefore, we distinguish between the \emph{\ac{IPFS} overlay} which entails all nodes, including clients, and the \emph{overlay without clients} which consists of nodes that take part in the \ac{DHT} protocol.
When no ambiguity may occur, we simply use the term overlay for both.

Node IDs and keys share the same representation and are treated identically as \emph{IDs} ($\in [0, 2^{256}]$).
Records in a Kademlia DHT are stored at nodes whose node ID is ``close'' to the record's key.
Closeness is defined via the bitwise XOR of two IDs, taken as an integer value, \ie, $\dist(x, y) = x \oplus y$.
A node stores its known neighbors in so-called \emph{$k$-buckets} which partition the known overlay network (without clients) based on the local node's ID.
Every $k$-bucket (or simply bucket) stores up to $k$ neighbors.
For each possible distance $n$ there is exactly one bucket storing nodes with $\dist(x, y) = n$, where $x$ is the local node's ID and $y$ is any node ID from the respective bucket.
In \ac{IPFS}' \ac{DHT}, $k = 20$, i.\,e., each bucket stores up to 20 entries.
Bucket $i$ stores nodes whose distance is in $[2^i, 2^{i+1})$, which effectively corresponds to the length of the common prefix shared between two node IDs.

\ac{IPFS} uses a performance optimization in that buckets are \emph{unfolded}, i.\,e., created, only when necessary.
As each bucket corresponds to a common prefix length, the ``later'' buckets with a longer common prefix length tend to be mostly empty, as the probability to encounter a node that falls into a specific bucket decreases exponentially with the common prefix length.
To avoid storing mostly empty buckets, \ac{IPFS} creates them on the fly when necessary.
This in turn implies that we can not know in advance how many buckets a node on the network has currently unfolded.

In \ac{IPFS}, a node $\node$ resolves an ID $x$ to a record or a set of closest nodes on the overlay using multi-round iterative \emph{lookups}.
In each lookup round, $\node$ sends requests to known nodes with IDs close to $x$.
Nodes reply with either a record corresponding to $x$ or with new node contacts, selected by proximity to $x$ and therefore enabling a subsequent lookup round.
Peer-to-peer connections in \ac{IPFS} are always based on TCP or other reliable transport layer protocols.
Consequently, the \ac{DHT} component of \ac{IPFS} also uses (TCP-) connections (unlike many other Kademlia implementations, which are datagram based).

In contrast to the original Kademlia proposal~\cite{DBLP:conf/iptps/MaymounkovM02}, \ac{IPFS} does not ping the last entry of a bucket for a potential eviction, if a new node is about to enter a new bucket.
Instead, the new node is simply rejected.
Nodes are ejected from buckets only if the connection with them is terminated for other reasons.
Similarly, \ac{IPFS}' \ac{DHT} does not maintain a replacement buffer~\cite{DBLP:conf/iptps/MaymounkovM02} with candidate nodes for full buckets.

\subsection{Content Retrieval: Bitswap Data Exchange}
\label{subsec:bitswap_exchange}

\emph{Bitswap} is a subprotocol of \ac{IPFS} used for transfers of actual content data.
It is similar to Bittorrent \cite{DBLP:conf/iptps/PouwelseGES05} and is used for obtaining data items from connected peers.
Interestingly, \ac{IPFS} overlay nodes (including clients) query \emph{all} of their connected peers for data items they look for (their so called \emph{wantlist}).
This happens in parallel to \ac{DHT} queries for data providers registered in the \ac{DHT}.
While in some aspects questionable with regards to performance, the promiscuous broadcasting of queries results in a key security feature of \ac{IPFS}:
attacks on the \ac{DHT} have only a limited impact on the availability of stored data.
For example, while it is within the means of a dedicated attacker to overwrite the providers record for a file with Sybil nodes, users are able to retrieve it via one of their many (around 900, cf. \cref{sec:results}) directly connected peers.

We verified this behavior experimentally within a private \ac{IPFS} network.
We overwrote the all provider lists for a target data item with Sybil entries,
effectively removing all correct providers from the \ac{DHT}.
Whereas, as expected, all \ac{DHT} queries for the data item failed after this intervention,
nodes were still able to obtain the data item from overlay neighbors.

\section{Understanding the Overlay Structure}
\label{sec:overlay_structure}

\ac{IPFS} nodes can be separated into two kinds: nodes that participate in the \ac{DHT} and nodes that are \emph{clients}.
We want to disentangle the different kinds of overlays that arise through this distinction and reason about what information can be measured.
We distinguish between the \ac{IPFS} overlay ($\tilde{G}$), the \ac{IPFS} overlay without clients ($G$) and the overlay induced by \ac{DHT} buckets ($G'$).
We explain the differences in the following.

Overlay networks are commonly modeled as graphs with nodes as vertices and connections between those nodes as edges.
Let $\tilde{G} = (\tilde{V}, \tilde{E})$ be an undirected graph (due to the symmetry of TCP-connections), with $\tilde{V}$ representing the nodes in the \ac{IPFS} overlay and $\tilde{E}$ the connections among them.
$\tilde{G}$ consists of \emph{all} nodes, including clients.
Since clients do not contribute to the \ac{DHT} -- one of the core pillars of \ac{IPFS} -- we focus most of our analysis on the \ac{IPFS} overlay without clients.
Let $V \subseteq \tilde{V}$ be the set of \ac{IPFS} nodes that are not clients and $G := \tilde{G}[V]$ the induced subgraph of $V$, \ie, the graph with vertex set $V$ and all edges from $\tilde{E}$ that have both endpoints in $V$.
$V$ can be, to a large extent, learned by crawling the \ac{DHT} buckets of each node, whereas it is not straightforward to reason about $\tilde{V}$ as client nodes are not registered anywhere or announced on the network.

What exactly can be learned by crawling each node's buckets?
\ac{IPFS} attempts to reflect every new connection (inbound or outbound) to \ac{DHT}-enabled nodes (i.\,e., nodes that are not pure clients) in \ac{DHT} buckets.
When connections are terminated for whatever reason, the corresponding entry is deleted from the respective bucket.
Each node's buckets, and hence the \ac{DHT} as a whole, therefore contains only active connections.
\emph{If} buckets were arbitrarily large, nearly all overlay links ($E$) would (also) be part of the buckets.

However, buckets are limited to $k = 20$ entries without an eviction mechanism.
This leads to situations where connections are established but cannot be reflected in buckets.
For example, during lookups, \ac{IPFS} establishes connections to all newly discovered nodes and attempts to add them to buckets.
To avoid maintaining a nearly infinite number of connections, \ac{IPFS} nodes start to randomly terminate connections that are older than \SI{30}{\second} once their total number of connections exceeds 900.
If relevant buckets are full at both nodes, the connection exists without being reflected in any bucket at all.
Let $G' = (V', E')$ be the bucket-induced subgraph of $G$,
i.\,e., $E' = \{ (v_i, v_j) \in E : v_j \text{ is in a bucket of } v_i \}$ and $V' \subset V$ is the set of all nodes used in $E'$.

\begin{figure}[htpb]
  \centering
\begin{tikzpicture}[node distance = 6em, inner sep=0.05cm]

	\node[draw, circle, color=black] (v0) at (0,0) {$v_0$};
	\node[draw, circle, color=black] (v1) at (3,0) {$v_1$};
	\node[draw, circle, color=black] (v2) at (1.5,1) {$v_2$};
	\node[draw, circle, color=black] (v3) at (4.5,1) {$v_3$};
	\node [left of=v0] (labelDHT) {$G$ (Overlay)};

		\path [<->] (v0) edge [draw] (v1);
		\path [<->] (v0) edge [draw] (v2);
		\path [<->] (v0) edge [draw] (v3);

		\path [<->] (v1) edge [draw] (v3);
		\path [<->] (v2) edge [draw] (v3);

		\node[draw, circle, color=blue, text=black] [above of=v0] (v0dht) {$v_0$};
	\node[draw, circle, color=blue, text=black] [above of=v1] (v1dht) {$v_1$};
	\node[draw, circle, color=blue, text=black] [above of=v2] (v2dht) {$v_2$};
	\node[draw, circle, color=blue, text=black] [above of=v3] (v3dht) {$v_3$};
	\node[text=blue] [left of=v0dht] (labelBucket) {$G'$ (Buckets)};

	\path (v0) edge [draw, dotted] (v0dht);
	\path (v1) edge [draw, dotted] (v1dht);
	\path (v2) edge [draw, dotted] (v2dht);
	\path (v3) edge [draw, dotted] (v3dht);

	\path[->] (v0dht) edge [draw, dashed, color=blue] (v1dht);
	\path[<-] (v0dht) edge [draw, dashed, color=blue] (v2dht);
	\path[<->] (v2dht) edge [draw, dashed, color=blue] (v3dht);
	\path[<->] (v1dht) edge [draw, dashed, color=blue] (v3dht);
\end{tikzpicture}
   \caption{Types of connections: non-client overlay (bottom) and buckets (top).}
  \label{fig:connection-layers}
\end{figure}
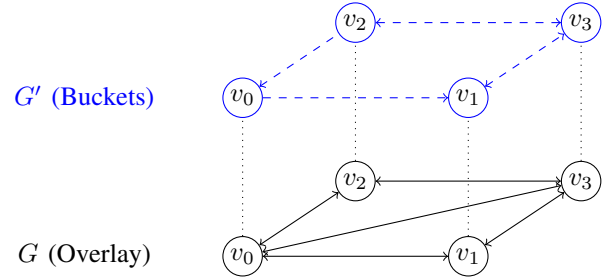

For visualizing the possible configurations, \cref{fig:connection-layers} depicts an example overlay network without clients ($G$), with the connections that are reflected in buckets ($G'$) redrawn on top.
A connection between two nodes can either
\begin{enumerate}
  \item be reflected in buckets by both nodes (e.\,g., $(v_2, v_3)$),
  \item be reflected in buckets by only one node (e.\,g., $(v_0, v_1)$) or
  \item exist independently of bucket entries (e.\,g., $(v_0, v_3)$).
\end{enumerate}
In practice, it therefore usually holds that $E'$ is a strict subset of $E$ and $G'$ therefore strict subgraph of $G$.
Unlike $G$, $G'$ is furthermore a \emph{directed} graph.
The direction is important since lookups of data or other nodes \emph{only} consider the nodes stored in buckets.

In the case of a well functioning \ac{DHT} it is expected that $V' \equiv V$ and $\tilde{V} \setminus V'$ is exactly the set of client nodes.
Event though it is likely that not all node pairs from $E$ are reflected in $E'$, we expect that $G'$ is strongly connected and that it is therefore, in principle, possible to enumerate all non-client nodes $V$ by crawling the \ac{DHT}.
Our crawling results (cf. \cref{subsec:num_nodes}) support this assumption.
Before delving into our crawling methodology and the remaining insights gained through it,
we present an empirically founded estimation of the relationship between $\tilde{G}$, $G$ and $G'$.

\section{Measuring the Interplay between $\tilde{G}$, $G$ and $G'$}
\label{sec:running_nodes}

In the following, we focus on the default behavior of nodes (with and without NAT), the share of clients in
the \ac{IPFS} overlay (\ie, $\tilde{V} \setminus V$) and how much of the overlay of \ac{DHT}-enabled nodes ($G$) is reflected in buckets ($G'$).

\subsection{Empirical Results of Monitoring nodes}
\label{subsec:emp_results_monitor}

First, we started an \ac{IPFS} node, version v0.5.0-dev with default settings and a public IP address, without any firewall.
Every \SI{20}{\second}, we queried the number of overlay connections, as well as the number of nodes in the node's buckets for a total of \SI{2.99}{\day}.
For the number of overlay connections we relied on the API that \ac{IPFS} offers in this case,
whereas the number of nodes was extracted through our external crawling tool (discussed in \cref{sec:crawler}).
\begin{figure}
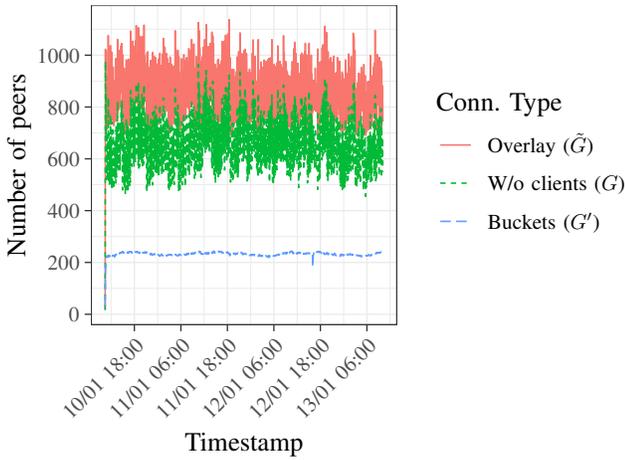


   \caption{Number of connections of an \ac{IPFS} node with default settings. We distinguish between all overlay connections, the overlay without clients and the buckets.}
  \label{fig:num_connected_peers_vs_dht}
\end{figure}
The results are depicted in \cref{fig:num_connected_peers_vs_dht}, which shows the number of overlay connections (edges from $\tilde{E}$, solid red line), connections with \ac{DHT}-enabled nodes (edges from $E$, dashed green line) and the number of connections in the buckets (edges from $E'$, dashed blue line).
The measurement began simultaneously with the \ac{IPFS} node itself, hence, the short start-up phase in the beginning.
It can be seen that the number of connections largely fluctuates around a value, with up to 400 connections established in just a few minutes.
This behavior is due to the way \ac{IPFS} handles its connection limit.

The default number of connections an \ac{IPFS} node will establish is 900, but it does not cap at that value.
Instead, \ac{IPFS} 1) starts a new connection with every node it encounters when querying data and 2) accepts every incoming connection at first.
If the number of connections exceeds the limit, \ac{IPFS} evicts connections (uniformly) at random that are older than \SI{30}{\second}, until the upper limit is satisfied again.
Furthermore, \SI{76.4}{\percent} of connections are \ac{DHT}-enabled, on average, indicating a notable difference between the overlay with clients ($\tilde{G}$) and the one without ($G$).
\begin{figure}
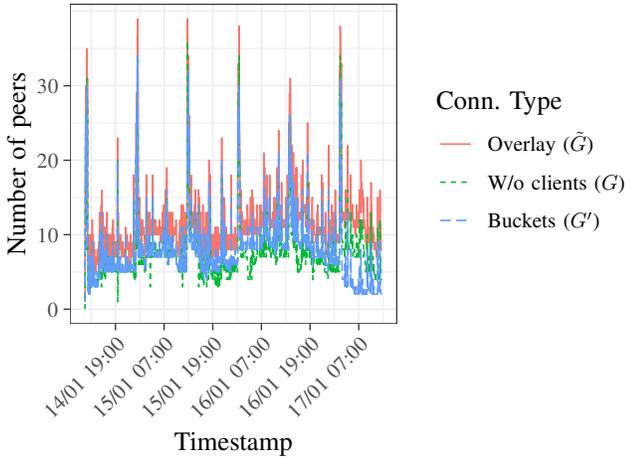


   \caption{The same as \cref{fig:num_connected_peers_vs_dht}, but behind a firewall.}
  \label{fig:nated_node}
\end{figure}
We performed the same experiment for a node behind a firewall for a duration of \SI{3.05}{\day}, the results are shown in \cref{fig:nated_node}.
Several properties can be observed from both experiments.
Firstly, a node behind a NAT has almost two orders of magnitude less connections than its non-NATed counterpart.
Secondly, most connections of the non-NATed node are \emph{inbound} connections, \ie, established from another peer on the network.
Since our node was idle and not searching for content or other peers, it has no external trigger to start outbound connections.
The NATed node cannot accept incoming connections, hence, the low number of connections.
Lastly, it can be seen that \ac{IPFS}' API functions lag behind in reporting the protocols of a peer: the number of \ac{DHT}-enabled connections can never be smaller than the number of connections in the buckets.
\begin{figure}
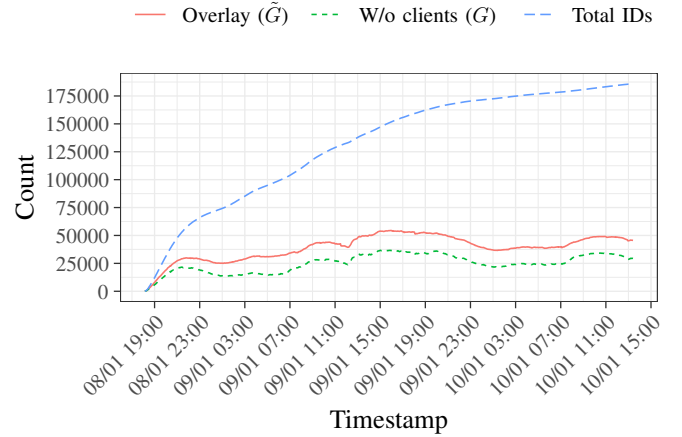


   \caption{Number of IDs seen, overlay connections and connections to \ac{DHT}-enabled peers of a node that has no connection limit.}
  \label{fig:no_limits_node}
\end{figure}
Last but not least, we are interested in a more holistic perspective on the different overlay types and number of nodes in the network.
To this end we started an \ac{IPFS} node (v0.5.0-dev) with a public IP and a connection limit of 500000.
Similar to the other experiments, we logged the connections every \SI{20}{\second} for a total of \SI{1.8}{\day}.
The results are depicted in \cref{fig:no_limits_node}, which shows the number of connections over time, the number of \ac{DHT} connections and the total number of node IDs seen.
On average, the node had 38903 connections, \SI{63.64}{\percent} of which were \ac{DHT} connections.
Again, the vast majority of these connections is inbound, since our node was idle.
The number of node IDs is steadily increasing, whereas the number of connections is not, which could be due to clients with non-persistent node IDs, \eg, the browser-enabled Javascript implementation of \ac{IPFS}\footnote{\url{https://github.com/ipfs/js-ipfs}}.

\subsection{Analysis of the Difference between $E$ and $E'$}
\label{subsec:quality_of_crawl}

In \cref{fig:num_connected_peers_vs_dht} it is clearly visible that the buckets ($E'$) store \SI{22.16}{\percent} of all connections between \ac{DHT}-enabled nodes ($E$), due to the buckets' limited capacity of $k=20$ nodes (cf. \cref{sec:overlay_structure}).
These connections can, therefore, \emph{not} be found by crawling the \ac{DHT} nor be obtained through passive measurements alone.
This raises the question: why is the gap between connections stored in a node's buckets ($E'$) and all overlay connections between non-client nodes ($E$) so significant?

Analytically, we are interested in the following quantity: Given $N := |V|$ non-client nodes in the overlay, what is the expected number of nodes stored in the buckets?
The distribution of nodes to buckets is highly skewed, since the first bucket is ``responsible'' for one half of the ID space, the second bucket for one fourth etc.~\cite{DBLP:conf/eurosp/HenningsenTF019,DBLP:journals/iacr/MarcusHG18}.

The expected number of nodes in each bucket $i$ is therefore $\min\{20, N\cdot p_i\}$, with $p_i := 2^{-i}$.
Although we do not encounter every peer equally likely, this reasoning still holds:
During bootstrap, an \ac{IPFS} node performs lookups for its own node ID as well as random targets tailored for each bucket.
The former ensures that it knows \emph{all} nodes in its direct neighborhood, partially filling the smallest, non-empty bucket.
The latter ensures that it knows some nodes from each other bucket, filling them completely with high probability.
Which bucket will be the smallest non-empty bucket therefore only depends on the total number of non-client nodes in the overlay.

\begin{figure}
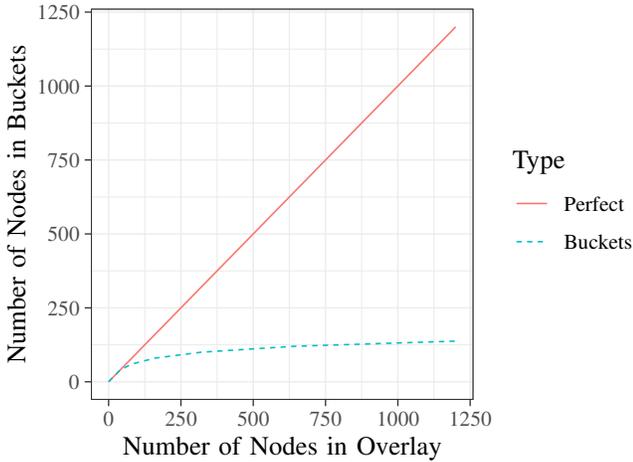


   \caption{Analytical approximation: number of overlay connections vs. the number of nodes actually stored in the buckets.}
  \label{fig:num_nodes_in_buckets}
\end{figure}

Abusing notation, this yields:
\begin{align}
  &\Ex[\text{\# nodes in buckets} | N \text{ nodes in overlay}]\\
  &= \sum_{i=1}^{256} \Ex[\text{\# nodes in bucket }i | N \text{ nodes in overlay}]\\
  &= \sum_{i=1}^{256} \min\{20, N\cdot p_i\}. \label{eq:nodes_in_buckets}
\end{align}
The result of this analytical consideration are depicted in \cref{fig:num_nodes_in_buckets}.
The solid red line corresponds to a ``perfect'' setting, where each overlay connection would be stored in a bucket, whereas the dashed blue line is the result of \cref{eq:nodes_in_buckets}.
Plugging the empirically found number of nodes from \cref{sec:results} into this formula yields an expected number of bucket entries between 232 and 247, which coincides with the measured average number of entries of 232.47 (max. 245).

So far, we've seen some indications on the relationship between the \ac{IPFS} overlay $\tilde{G}$, the overlay without clients $G$ and what is stored in buckets $G'$.
In the following, we are interested in obtaining measurements of $G'$ to learn more about the topology of the network.

\section{Crawling the Kademlia \ac{DHT}}
\label{sec:crawler}

\subsection{Crawl Procedure}

As described in \cref{sec:ipfs-nutshell}, nodes in the network are identified through a (multi-)hash of a public key.
The crawler used new key pairs in every run to thwart potential biases due to re-occurring IDs.
Moreover, an \ac{IPFS} node will not store our crawling nodes in its buckets, since our nodes are marked as clients who do not actively participate in the Kademlia communication\footnote{Nodes exchange a list of protocols they can serve. By not including the DHT protocol in said list, other nodes know that we cannot answer DHT messages.}.

Nodes can be found by issuing FindNode packets for some target ID, which can be a key or a node ID (=hash of a public key in the case of nodes).
To completely crawl a node, one has to send a FindNode packet for each possible bucket.
This is due to the fact that a node returns its $k$ closest neighbors to the target provided in a FindNode packet.
The closest neighbors to the target are the ones in the bucket the target falls into.
If the bucket in question is not full (\ie, less than $k$ entries), the closest neighbors are the ones in the target bucket and the two buckets surrounding it.
Since the bucket utilization of remote nodes is unknown,
we do not know in advance how many requests per node we have to send for obtaining all of its \ac{DHT} neighbors.
We therefore send FindNode packets to targets with increasing common prefix lengths and stop once no new nodes are learned.
This results in a significant speed improvement as no requests are sent for buckets that are nearly certainly empty
(since the number of potential bucket entries decreases exponentially with the common prefix length).
Faster crawls, in turn, enable us to capture more accurate snapshots of the dynamically changing network.

\subsection{Hash Pre-Image Computation}

Unlike more canonical Kademlia implementations, whenever an \ac{IPFS} node receives a FindNode packet, it hashes the provided target and searches for the nearest neighbors to that hash.
Therefore, to know which target to send for which bucket, we need to compute pre-images that, when hashed, yield the desired common prefix length between FindNode target and the node ID of the node we are crawling.
To that end, we generated pre-images for every possible prefix of 24-bit length.
In other words, we computed $2^{24}$ pre-images such that, for each possible prefix of the form $\{0, 1\}^{24}$, there is a hash starting with that prefix.

Equipped with this table, one lookup operation is sufficient to pick a pre-image that, when hashed by the receiver, will yield the desired bucket.
Note that this table can be used for an arbitrary number of crawls, hence, the computation only had to be performed \emph{once}.

\subsection{Crawl Procedure}

Each crawl commenced with the 4 IPFS default bootnodes
and continued from there by sending FindNode packets for each common prefix length (and therefore bucket) until no more new nodes were learned.
Therefore, if the IPFS network were static and nodes replied to requests deterministically, a crawl would always yield the same results.
Due to the inherent dynamics in the overlay this is not the case: repeated crawls allow us to observe changes in the overlay over time.

\section{Crawling Results}
\label{sec:results}

We repeatedly crawled the IPFS network for a total duration of \SI{6.98}{\day}.
Crawls were started one at a time and back to back, in the sense that as soon as a crawl was finished, a new one was started.
We performed a total of 2400 crawls, with a single crawl taking an average of \SI{4.1}{\minute} to complete.

\subsection{Number of Nodes, Reachability and Churn}
\label{subsec:num_nodes}

To get an idea for the size of the network, we first focus on the number of nodes in the network and their session lengths.
During our \SI{6.98}{\day}, we found a total of 309404 distinct node IDs,
with an average number of 44474 per crawl.
This is consistent with the results obtained in \cref{sec:running_nodes},
hinting that both methods provide an adequate view of $V$, the set of non-client nodes in the \ac{IPFS} overlay.
Surprisingly, of all the nodes that were queried, the crawler was only able to connect to \SI{6.55}{\percent}, on average.

We suspect that most \ac{IPFS} nodes are run by private people connected through NAT.
This hypothesis is supported by our results: about \SI{52.19}{\percent} of all nodes report only local IP addresses for other nodes to connect to, which is exactly the behavior of nodes behind symmetric NATs (cf. \cref{subsec:eval_country_distribution}).
Furthermore, if most nodes are behind NATs, they are also prone to short uptimes, since these are probably used as client nodes which are shut down after use.

Exactly this behavior can be observed regarding the session lengths, which are depicted in \cref{tab:sessionLengths}.
We define a session as the time difference between the crawl, when we were able to reach the node and when it became unreachable again.
The table depicts the inverse cumulative session lengths: each row yields the number of sessions (and their percentage) that were longer than the given duration.
For example, roughly \SI{56}{\percent} of all sessions were longer than \SI{5}{\minute}, or equivalently, \SI{44}{\percent} of all sessions were \emph{shorter} than \SI{5}{\minute}.
\begin{table}
  \center
\begin{tabular}{| c | c | c |}
 \hline
 Session Duration & Percentage & Number of sessions\\
 \hline
5 minutes & 56.9273 & 2188153 \\
 \hline
10 minutes & 26.1066 & 1003478 \\
 \hline
30 minutes & 2.6809 & 103049 \\
 \hline
1 hour & 0.4289 & 16486 \\
 \hline
1 day & 8e-04 & 32 \\
 \hline
6 days & 1e-04 & 2 \\
 \hline
\end{tabular}
   \caption{Inverse cumulative session lengths: each row gives the number of sessions (and total percentage) that were \emph{longer} than the given duration.}
  \label{tab:sessionLengths}
\end{table}
This indicates that participation in \ac{IPFS} is dynamic and rather short-lived.
\begin{figure}
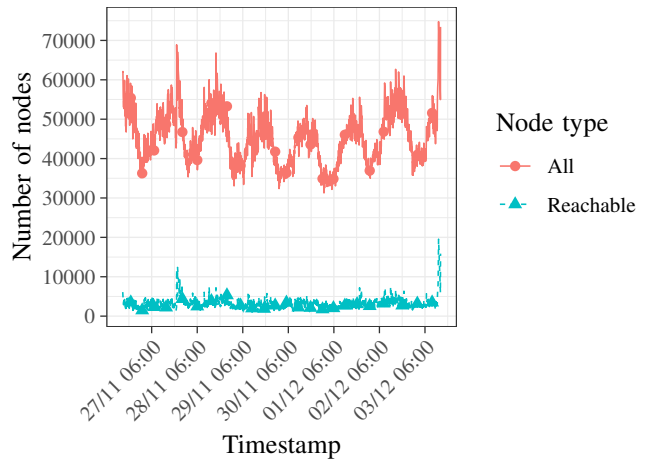

%
   \caption{Number of nodes over time, distinguished by all and reachable (=answered to our query) nodes. Times are in UTC.}
  \label{fig:num_nodes}
\end{figure}
We also observed a periodic pattern in the number of nodes found through crawling, as shown in \cref{fig:num_nodes}.
The figure distinguishes between all nodes and nodes that were reachable, i.\,e., the crawler was able to establish a connection to these nodes.
The significant increase in the number of nodes at the end of the measurement period could stem from other researchers' nodes or even an attack on the network.
It can be seen that between noon and 19pm UTC,
the number of nodes increases significantly.
This might hint at a usage pattern in that users start their \ac{IPFS} nodes on-demand in their spare time and shut them down after use.

Additionally, this underlines the hypothesis of most nodes being operated by private people behind NATs, as the number of \emph{reachable} nodes, i.\,e., nodes we could connect to, does not fluctuate as much.

One could argue that private users behind NATs will tend to use their nodes in their spare time in the evening.
Hence, a usage pattern with peaks in the afternoon of UTC time hints to many users in Asia, since UTC afternoon corresponds to evening times in, e.\,g., China.
This hypothesis is supported by the distribution of nodes over countries.

\subsection{Node Distribution over Countries and Protocol Usage} 
\label{subsec:eval_country_distribution}
\begin{table}[htb]
\begin{tabular}{| c | c | c || c | c | c |}
 \hline
 \multicolumn{3}{| c ||}{All} & \multicolumn{3}{| c |}{Reachable}\\
 \hline
 Country & Count & Conf. Int. & Country & Count & Conf. Int.\\
 \hline
LocalIP & 23973.61 & $\pm$173.85 & US & 1721.08 & $\pm$26.67 \\
 \hline
CN & 5631.37 & $\pm$11.27 & FR & 635.71 & $\pm$17.65 \\
 \hline
DE & 4224.76 & $\pm$33.37 & DE & 569.23 & $\pm$8.82 \\
 \hline
US & 4091.64 & $\pm$54.13 & CA & 119.01 & $\pm$2.32 \\
 \hline
FR & 1390.2 & $\pm$33.33 & PL & 73.63 & $\pm$2.09 \\
 \hline
CA & 978.17 & $\pm$19.87 & CN & 58.61 & $\pm$0.49 \\
 \hline
PL & 693.08 & $\pm$17.97 & GB & 26.67 & $\pm$0.23 \\
 \hline
IL & 321.07 & $\pm$2.63 & SG & 20.98 & $\pm$0.15 \\
 \hline
GB & 171.08 & $\pm$0.91 & IL & 15.09 & $\pm$0.35 \\
 \hline
HK & 168.09 & $\pm$1.96 & NL & 12.95 & $\pm$0.12 \\
 \hline
\end{tabular}
 \caption{The top ten countries per crawl, differentiated by all discovered nodes and nodes that were reachable. Depicted is the average count per country per crawl as well as confidence intervals.}
\label{tab:geoip}
\end{table}
\begin{table}[htb]
  \center
\begin{tabular}{| c | c | c |}
 \hline
 Protocol & Perc. of peers & Abs. count\\
 \hline
ip4 & 99.9893 & 309369 \\
 \hline
ip6 & 80.4862 & 249026 \\
 \hline
p2p-circuit & 1.5313 & 4738 \\
 \hline
ipfs & 1.2737 & 3941 \\
 \hline
dns4 & 0.0669 & 207 \\
 \hline
dns6 & 0.0301 & 93 \\
 \hline
dnsaddr & 0.0039 & 12 \\
 \hline
onion3 & 3e-04 & 1 \\
 \hline
\end{tabular}
   \caption{Protocol Usage.}
  \label{tab:protocols}
\end{table}
\cref{tab:geoip} depicts the top ten countries, both for all discovered nodes and for nodes that were reachable by our crawler\footnote{
  We use the GeoLite2 data from MaxMind for this task, available at \url{https://www.maxmind.com}.}
These ten countries contain \SI{91.04}{\percent} (\SI{93.67}{\percent} in the case of reachable nodes) of the whole network, already hinting at centralization tendencies regarding the spatial distribution of nodes.

Again, it can be seen that \SI{52.19}{\percent} of all nodes \emph{only} provide local or private IP addresses, thus making it impossible to connect to them.
This is in line with the default behavior of \ac{IPFS} when operated behind a NAT.
When a node first comes online, it does not know its external IP address and therefore advertises the internal IP addresses to peers it connects to.
These entries enter the \ac{DHT}, since \ac{IPFS} aims to provide support for building private \ac{IPFS} networks.
Over time, a node learns about its external multi-addresses (multi-addresses contain address and port, cf. \cref{sec:ipfs-nutshell}) from its peers.
An \ac{IPFS} considers these observed multi-addresses reachable, if at least four peers have reported the same multi-address in the last 40 minutes and the multi-address has been seen in the last ten minutes.
This is never the case for symmetric NATs, which assign a unique external address and port for every connection, yielding a different multi-address for every connected peer.
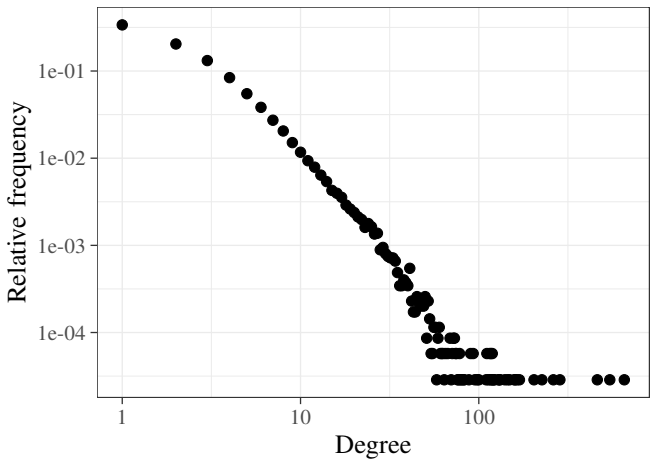
\begin{figure}[!htb]
\centering
\begin{tikzpicture}[x=1pt,y=1pt]
\definecolor{fillColor}{RGB}{255,255,255}
\path[use as bounding box,fill=fillColor,fill opacity=0.00] (0,0) rectangle (252.94,180.67);
\begin{scope}
\path[clip] (  0.00,  0.00) rectangle (252.94,180.67);
\definecolor{drawColor}{RGB}{255,255,255}
\definecolor{fillColor}{RGB}{255,255,255}

\path[draw=drawColor,line width= 0.5pt,line join=round,line cap=round,fill=fillColor] ( -0.00,  0.00) rectangle (252.95,180.68);
\end{scope}
\begin{scope}
\path[clip] ( 39.05, 27.90) rectangle (247.95,175.68);
\definecolor{fillColor}{RGB}{255,255,255}

\path[fill=fillColor] ( 39.05, 27.90) rectangle (247.95,175.68);
\definecolor{drawColor}{gray}{0.92}

\path[draw=drawColor,line width= 0.3pt,line join=round] ( 39.05, 36.01) --
	(247.95, 36.01);

\path[draw=drawColor,line width= 0.3pt,line join=round] ( 39.05, 69.01) --
	(247.95, 69.01);

\path[draw=drawColor,line width= 0.3pt,line join=round] ( 39.05,102.00) --
	(247.95,102.00);

\path[draw=drawColor,line width= 0.3pt,line join=round] ( 39.05,135.00) --
	(247.95,135.00);

\path[draw=drawColor,line width= 0.3pt,line join=round] ( 39.05,168.00) --
	(247.95,168.00);

\path[draw=drawColor,line width= 0.3pt,line join=round] ( 82.26, 27.90) --
	( 82.26,175.68);

\path[draw=drawColor,line width= 0.3pt,line join=round] (149.69, 27.90) --
	(149.69,175.68);

\path[draw=drawColor,line width= 0.3pt,line join=round] (217.12, 27.90) --
	(217.12,175.68);

\path[draw=drawColor,line width= 0.5pt,line join=round] ( 39.05, 52.51) --
	(247.95, 52.51);

\path[draw=drawColor,line width= 0.5pt,line join=round] ( 39.05, 85.50) --
	(247.95, 85.50);

\path[draw=drawColor,line width= 0.5pt,line join=round] ( 39.05,118.50) --
	(247.95,118.50);

\path[draw=drawColor,line width= 0.5pt,line join=round] ( 39.05,151.50) --
	(247.95,151.50);

\path[draw=drawColor,line width= 0.5pt,line join=round] ( 48.54, 27.90) --
	( 48.54,175.68);

\path[draw=drawColor,line width= 0.5pt,line join=round] (115.98, 27.90) --
	(115.98,175.68);

\path[draw=drawColor,line width= 0.5pt,line join=round] (183.41, 27.90) --
	(183.41,175.68);
\definecolor{drawColor}{RGB}{0,0,0}
\definecolor{fillColor}{RGB}{0,0,0}

\path[draw=drawColor,line width= 0.4pt,line join=round,line cap=round,fill=fillColor] ( 48.54,168.96) circle (  1.96);

\path[draw=drawColor,line width= 0.4pt,line join=round,line cap=round,fill=fillColor] ( 68.84,161.73) circle (  1.96);

\path[draw=drawColor,line width= 0.4pt,line join=round,line cap=round,fill=fillColor] ( 80.72,155.45) circle (  1.96);

\path[draw=drawColor,line width= 0.4pt,line join=round,line cap=round,fill=fillColor] ( 89.14,148.99) circle (  1.96);

\path[draw=drawColor,line width= 0.4pt,line join=round,line cap=round,fill=fillColor] ( 95.68,142.90) circle (  1.96);

\path[draw=drawColor,line width= 0.4pt,line join=round,line cap=round,fill=fillColor] (101.02,137.75) circle (  1.96);

\path[draw=drawColor,line width= 0.4pt,line join=round,line cap=round,fill=fillColor] (105.53,132.87) circle (  1.96);

\path[draw=drawColor,line width= 0.4pt,line join=round,line cap=round,fill=fillColor] (109.44,128.80) circle (  1.96);

\path[draw=drawColor,line width= 0.4pt,line join=round,line cap=round,fill=fillColor] (112.89,124.40) circle (  1.96);

\path[draw=drawColor,line width= 0.4pt,line join=round,line cap=round,fill=fillColor] (115.98,120.76) circle (  1.96);

\path[draw=drawColor,line width= 0.4pt,line join=round,line cap=round,fill=fillColor] (118.77,117.54) circle (  1.96);

\path[draw=drawColor,line width= 0.4pt,line join=round,line cap=round,fill=fillColor] (121.32,115.10) circle (  1.96);

\path[draw=drawColor,line width= 0.4pt,line join=round,line cap=round,fill=fillColor] (123.66,112.10) circle (  1.96);

\path[draw=drawColor,line width= 0.4pt,line join=round,line cap=round,fill=fillColor] (125.83,109.65) circle (  1.96);

\path[draw=drawColor,line width= 0.4pt,line join=round,line cap=round,fill=fillColor] (127.85,106.32) circle (  1.96);

\path[draw=drawColor,line width= 0.4pt,line join=round,line cap=round,fill=fillColor] (129.74,105.22) circle (  1.96);

\path[draw=drawColor,line width= 0.4pt,line join=round,line cap=round,fill=fillColor] (131.52,103.69) circle (  1.96);

\path[draw=drawColor,line width= 0.4pt,line join=round,line cap=round,fill=fillColor] (133.19,100.75) circle (  1.96);

\path[draw=drawColor,line width= 0.4pt,line join=round,line cap=round,fill=fillColor] (134.77, 99.26) circle (  1.96);

\path[draw=drawColor,line width= 0.4pt,line join=round,line cap=round,fill=fillColor] (136.28, 97.94) circle (  1.96);

\path[draw=drawColor,line width= 0.4pt,line join=round,line cap=round,fill=fillColor] (137.70, 96.29) circle (  1.96);

\path[draw=drawColor,line width= 0.4pt,line join=round,line cap=round,fill=fillColor] (139.07, 95.29) circle (  1.96);

\path[draw=drawColor,line width= 0.4pt,line join=round,line cap=round,fill=fillColor] (140.37, 92.30) circle (  1.96);

\path[draw=drawColor,line width= 0.4pt,line join=round,line cap=round,fill=fillColor] (141.62, 93.76) circle (  1.96);

\path[draw=drawColor,line width= 0.4pt,line join=round,line cap=round,fill=fillColor] (142.81, 92.55) circle (  1.96);

\path[draw=drawColor,line width= 0.4pt,line join=round,line cap=round,fill=fillColor] (143.96, 89.79) circle (  1.96);

\path[draw=drawColor,line width= 0.4pt,line join=round,line cap=round,fill=fillColor] (145.06, 90.09) circle (  1.96);

\path[draw=drawColor,line width= 0.4pt,line join=round,line cap=round,fill=fillColor] (146.13, 83.82) circle (  1.96);

\path[draw=drawColor,line width= 0.4pt,line join=round,line cap=round,fill=fillColor] (147.16, 84.72) circle (  1.96);

\path[draw=drawColor,line width= 0.4pt,line join=round,line cap=round,fill=fillColor] (148.15, 82.36) circle (  1.96);

\path[draw=drawColor,line width= 0.4pt,line join=round,line cap=round,fill=fillColor] (149.11, 81.30) circle (  1.96);

\path[draw=drawColor,line width= 0.4pt,line join=round,line cap=round,fill=fillColor] (150.04, 80.74) circle (  1.96);

\path[draw=drawColor,line width= 0.4pt,line join=round,line cap=round,fill=fillColor] (150.94, 80.74) circle (  1.96);

\path[draw=drawColor,line width= 0.4pt,line join=round,line cap=round,fill=fillColor] (151.82, 79.55) circle (  1.96);

\path[draw=drawColor,line width= 0.4pt,line join=round,line cap=round,fill=fillColor] (152.66, 75.21) circle (  1.96);

\path[draw=drawColor,line width= 0.4pt,line join=round,line cap=round,fill=fillColor] (153.49, 70.22) circle (  1.96);

\path[draw=drawColor,line width= 0.4pt,line join=round,line cap=round,fill=fillColor] (154.29, 70.22) circle (  1.96);

\path[draw=drawColor,line width= 0.4pt,line join=round,line cap=round,fill=fillColor] (155.07, 72.43) circle (  1.96);

\path[draw=drawColor,line width= 0.4pt,line join=round,line cap=round,fill=fillColor] (155.83, 71.37) circle (  1.96);

\path[draw=drawColor,line width= 0.4pt,line join=round,line cap=round,fill=fillColor] (156.57, 70.22) circle (  1.96);

\path[draw=drawColor,line width= 0.4pt,line join=round,line cap=round,fill=fillColor] (157.30, 76.81) circle (  1.96);

\path[draw=drawColor,line width= 0.4pt,line join=round,line cap=round,fill=fillColor] (158.00, 64.41) circle (  1.96);

\path[draw=drawColor,line width= 0.4pt,line join=round,line cap=round,fill=fillColor] (158.69, 60.29) circle (  1.96);

\path[draw=drawColor,line width= 0.4pt,line join=round,line cap=round,fill=fillColor] (159.37, 60.29) circle (  1.96);

\path[draw=drawColor,line width= 0.4pt,line join=round,line cap=round,fill=fillColor] (160.02, 66.10) circle (  1.96);

\path[draw=drawColor,line width= 0.4pt,line join=round,line cap=round,fill=fillColor] (160.67, 64.41) circle (  1.96);

\path[draw=drawColor,line width= 0.4pt,line join=round,line cap=round,fill=fillColor] (161.30, 62.50) circle (  1.96);

\path[draw=drawColor,line width= 0.4pt,line join=round,line cap=round,fill=fillColor] (161.91, 64.41) circle (  1.96);

\path[draw=drawColor,line width= 0.4pt,line join=round,line cap=round,fill=fillColor] (162.52, 62.50) circle (  1.96);

\path[draw=drawColor,line width= 0.4pt,line join=round,line cap=round,fill=fillColor] (163.11, 66.10) circle (  1.96);

\path[draw=drawColor,line width= 0.4pt,line join=round,line cap=round,fill=fillColor] (163.69, 50.36) circle (  1.96);

\path[draw=drawColor,line width= 0.4pt,line join=round,line cap=round,fill=fillColor] (164.26, 64.41) circle (  1.96);

\path[draw=drawColor,line width= 0.4pt,line join=round,line cap=round,fill=fillColor] (164.82, 57.68) circle (  1.96);

\path[draw=drawColor,line width= 0.4pt,line join=round,line cap=round,fill=fillColor] (165.36, 44.55) circle (  1.96);

\path[draw=drawColor,line width= 0.4pt,line join=round,line cap=round,fill=fillColor] (165.90, 44.55) circle (  1.96);

\path[draw=drawColor,line width= 0.4pt,line join=round,line cap=round,fill=fillColor] (166.43, 54.48) circle (  1.96);

\path[draw=drawColor,line width= 0.4pt,line join=round,line cap=round,fill=fillColor] (166.95, 54.48) circle (  1.96);

\path[draw=drawColor,line width= 0.4pt,line join=round,line cap=round,fill=fillColor] (167.46, 34.61) circle (  1.96);

\path[draw=drawColor,line width= 0.4pt,line join=round,line cap=round,fill=fillColor] (167.96, 50.36) circle (  1.96);

\path[draw=drawColor,line width= 0.4pt,line join=round,line cap=round,fill=fillColor] (168.45, 54.48) circle (  1.96);

\path[draw=drawColor,line width= 0.4pt,line join=round,line cap=round,fill=fillColor] (168.93, 44.55) circle (  1.96);

\path[draw=drawColor,line width= 0.4pt,line join=round,line cap=round,fill=fillColor] (169.41, 44.55) circle (  1.96);

\path[draw=drawColor,line width= 0.4pt,line join=round,line cap=round,fill=fillColor] (169.88, 44.55) circle (  1.96);

\path[draw=drawColor,line width= 0.4pt,line join=round,line cap=round,fill=fillColor] (170.34, 34.61) circle (  1.96);

\path[draw=drawColor,line width= 0.4pt,line join=round,line cap=round,fill=fillColor] (170.79, 44.55) circle (  1.96);

\path[draw=drawColor,line width= 0.4pt,line join=round,line cap=round,fill=fillColor] (171.68, 44.55) circle (  1.96);

\path[draw=drawColor,line width= 0.4pt,line join=round,line cap=round,fill=fillColor] (172.54, 50.36) circle (  1.96);

\path[draw=drawColor,line width= 0.4pt,line join=round,line cap=round,fill=fillColor] (172.96, 34.61) circle (  1.96);

\path[draw=drawColor,line width= 0.4pt,line join=round,line cap=round,fill=fillColor] (173.38, 44.55) circle (  1.96);

\path[draw=drawColor,line width= 0.4pt,line join=round,line cap=round,fill=fillColor] (173.79, 50.36) circle (  1.96);

\path[draw=drawColor,line width= 0.4pt,line join=round,line cap=round,fill=fillColor] (174.19, 50.36) circle (  1.96);

\path[draw=drawColor,line width= 0.4pt,line join=round,line cap=round,fill=fillColor] (174.59, 44.55) circle (  1.96);

\path[draw=drawColor,line width= 0.4pt,line join=round,line cap=round,fill=fillColor] (174.98, 44.55) circle (  1.96);

\path[draw=drawColor,line width= 0.4pt,line join=round,line cap=round,fill=fillColor] (175.37, 34.61) circle (  1.96);

\path[draw=drawColor,line width= 0.4pt,line join=round,line cap=round,fill=fillColor] (175.75, 34.61) circle (  1.96);

\path[draw=drawColor,line width= 0.4pt,line join=round,line cap=round,fill=fillColor] (176.13, 44.55) circle (  1.96);

\path[draw=drawColor,line width= 0.4pt,line join=round,line cap=round,fill=fillColor] (176.51, 34.61) circle (  1.96);

\path[draw=drawColor,line width= 0.4pt,line join=round,line cap=round,fill=fillColor] (176.87, 34.61) circle (  1.96);

\path[draw=drawColor,line width= 0.4pt,line join=round,line cap=round,fill=fillColor] (177.60, 34.61) circle (  1.96);

\path[draw=drawColor,line width= 0.4pt,line join=round,line cap=round,fill=fillColor] (177.95, 34.61) circle (  1.96);

\path[draw=drawColor,line width= 0.4pt,line join=round,line cap=round,fill=fillColor] (179.67, 34.61) circle (  1.96);

\path[draw=drawColor,line width= 0.4pt,line join=round,line cap=round,fill=fillColor] (180.32, 44.55) circle (  1.96);

\path[draw=drawColor,line width= 0.4pt,line join=round,line cap=round,fill=fillColor] (181.28, 44.55) circle (  1.96);

\path[draw=drawColor,line width= 0.4pt,line join=round,line cap=round,fill=fillColor] (181.91, 34.61) circle (  1.96);

\path[draw=drawColor,line width= 0.4pt,line join=round,line cap=round,fill=fillColor] (182.82, 34.61) circle (  1.96);

\path[draw=drawColor,line width= 0.4pt,line join=round,line cap=round,fill=fillColor] (183.41, 34.61) circle (  1.96);

\path[draw=drawColor,line width= 0.4pt,line join=round,line cap=round,fill=fillColor] (186.20, 34.61) circle (  1.96);

\path[draw=drawColor,line width= 0.4pt,line join=round,line cap=round,fill=fillColor] (186.46, 44.55) circle (  1.96);

\path[draw=drawColor,line width= 0.4pt,line join=round,line cap=round,fill=fillColor] (186.99, 34.61) circle (  1.96);

\path[draw=drawColor,line width= 0.4pt,line join=round,line cap=round,fill=fillColor] (187.50, 44.55) circle (  1.96);

\path[draw=drawColor,line width= 0.4pt,line join=round,line cap=round,fill=fillColor] (187.76, 34.61) circle (  1.96);

\path[draw=drawColor,line width= 0.4pt,line join=round,line cap=round,fill=fillColor] (188.01, 44.55) circle (  1.96);

\path[draw=drawColor,line width= 0.4pt,line join=round,line cap=round,fill=fillColor] (188.26, 34.61) circle (  1.96);

\path[draw=drawColor,line width= 0.4pt,line join=round,line cap=round,fill=fillColor] (188.50, 44.55) circle (  1.96);

\path[draw=drawColor,line width= 0.4pt,line join=round,line cap=round,fill=fillColor] (188.75, 34.61) circle (  1.96);

\path[draw=drawColor,line width= 0.4pt,line join=round,line cap=round,fill=fillColor] (188.99, 34.61) circle (  1.96);

\path[draw=drawColor,line width= 0.4pt,line join=round,line cap=round,fill=fillColor] (189.47, 34.61) circle (  1.96);

\path[draw=drawColor,line width= 0.4pt,line join=round,line cap=round,fill=fillColor] (190.87, 34.61) circle (  1.96);

\path[draw=drawColor,line width= 0.4pt,line join=round,line cap=round,fill=fillColor] (191.09, 34.61) circle (  1.96);

\path[draw=drawColor,line width= 0.4pt,line join=round,line cap=round,fill=fillColor] (191.54, 34.61) circle (  1.96);

\path[draw=drawColor,line width= 0.4pt,line join=round,line cap=round,fill=fillColor] (193.26, 34.61) circle (  1.96);

\path[draw=drawColor,line width= 0.4pt,line join=round,line cap=round,fill=fillColor] (194.29, 34.61) circle (  1.96);

\path[draw=drawColor,line width= 0.4pt,line join=round,line cap=round,fill=fillColor] (195.09, 34.61) circle (  1.96);

\path[draw=drawColor,line width= 0.4pt,line join=round,line cap=round,fill=fillColor] (196.80, 34.61) circle (  1.96);

\path[draw=drawColor,line width= 0.4pt,line join=round,line cap=round,fill=fillColor] (197.17, 34.61) circle (  1.96);

\path[draw=drawColor,line width= 0.4pt,line join=round,line cap=round,fill=fillColor] (197.90, 34.61) circle (  1.96);

\path[draw=drawColor,line width= 0.4pt,line join=round,line cap=round,fill=fillColor] (198.78, 34.61) circle (  1.96);

\path[draw=drawColor,line width= 0.4pt,line join=round,line cap=round,fill=fillColor] (204.29, 34.61) circle (  1.96);

\path[draw=drawColor,line width= 0.4pt,line join=round,line cap=round,fill=fillColor] (207.29, 34.61) circle (  1.96);

\path[draw=drawColor,line width= 0.4pt,line join=round,line cap=round,fill=fillColor] (211.62, 34.61) circle (  1.96);

\path[draw=drawColor,line width= 0.4pt,line join=round,line cap=round,fill=fillColor] (214.08, 34.61) circle (  1.96);

\path[draw=drawColor,line width= 0.4pt,line join=round,line cap=round,fill=fillColor] (228.23, 34.61) circle (  1.96);

\path[draw=drawColor,line width= 0.4pt,line join=round,line cap=round,fill=fillColor] (232.96, 34.61) circle (  1.96);

\path[draw=drawColor,line width= 0.4pt,line join=round,line cap=round,fill=fillColor] (238.45, 34.61) circle (  1.96);
\definecolor{drawColor}{gray}{0.20}

\path[draw=drawColor,line width= 0.5pt,line join=round,line cap=round] ( 39.05, 27.90) rectangle (247.95,175.68);
\end{scope}
\begin{scope}
\path[clip] (  0.00,  0.00) rectangle (252.94,180.67);
\definecolor{drawColor}{gray}{0.30}

\node[text=drawColor,anchor=base east,inner sep=0pt, outer sep=0pt, scale=  0.80] at ( 34.55, 49.75) {1e-04};

\node[text=drawColor,anchor=base east,inner sep=0pt, outer sep=0pt, scale=  0.80] at ( 34.55, 82.75) {1e-03};

\node[text=drawColor,anchor=base east,inner sep=0pt, outer sep=0pt, scale=  0.80] at ( 34.55,115.75) {1e-02};

\node[text=drawColor,anchor=base east,inner sep=0pt, outer sep=0pt, scale=  0.80] at ( 34.55,148.74) {1e-01};
\end{scope}
\begin{scope}
\path[clip] (  0.00,  0.00) rectangle (252.94,180.67);
\definecolor{drawColor}{gray}{0.20}

\path[draw=drawColor,line width= 0.5pt,line join=round] ( 36.55, 52.51) --
	( 39.05, 52.51);

\path[draw=drawColor,line width= 0.5pt,line join=round] ( 36.55, 85.50) --
	( 39.05, 85.50);

\path[draw=drawColor,line width= 0.5pt,line join=round] ( 36.55,118.50) --
	( 39.05,118.50);

\path[draw=drawColor,line width= 0.5pt,line join=round] ( 36.55,151.50) --
	( 39.05,151.50);
\end{scope}
\begin{scope}
\path[clip] (  0.00,  0.00) rectangle (252.94,180.67);
\definecolor{drawColor}{gray}{0.20}

\path[draw=drawColor,line width= 0.5pt,line join=round] ( 48.54, 25.40) --
	( 48.54, 27.90);

\path[draw=drawColor,line width= 0.5pt,line join=round] (115.98, 25.40) --
	(115.98, 27.90);

\path[draw=drawColor,line width= 0.5pt,line join=round] (183.41, 25.40) --
	(183.41, 27.90);
\end{scope}
\begin{scope}
\path[clip] (  0.00,  0.00) rectangle (252.94,180.67);
\definecolor{drawColor}{gray}{0.30}

\node[text=drawColor,anchor=base,inner sep=0pt, outer sep=0pt, scale=  0.80] at ( 48.54, 17.89) {1};

\node[text=drawColor,anchor=base,inner sep=0pt, outer sep=0pt, scale=  0.80] at (115.98, 17.89) {10};

\node[text=drawColor,anchor=base,inner sep=0pt, outer sep=0pt, scale=  0.80] at (183.41, 17.89) {100};
\end{scope}
\begin{scope}
\path[clip] (  0.00,  0.00) rectangle (252.94,180.67);
\definecolor{drawColor}{RGB}{0,0,0}

\node[text=drawColor,anchor=base,inner sep=0pt, outer sep=0pt, scale=  1.00] at (143.50,  6.94) {Degree};
\end{scope}
\begin{scope}
\path[clip] (  0.00,  0.00) rectangle (252.94,180.67);
\definecolor{drawColor}{RGB}{0,0,0}

\node[text=drawColor,rotate= 90.00,anchor=base,inner sep=0pt, outer sep=0pt, scale=  1.00] at ( 11.89,101.79) {Relative frequency};
\end{scope}
\end{tikzpicture}
         \caption{In-Degree distribution from the first crawl including \emph{all} found nodes. Other crawls yielded similar shapes.}
        \label{fig:log_all_nodes_degree_distribution}
\end{figure}
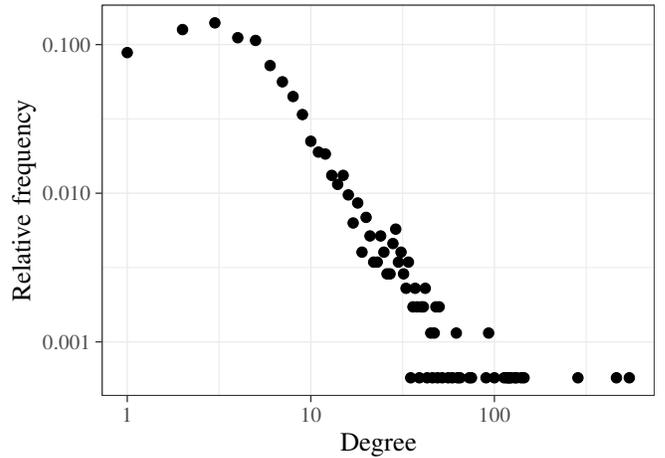
\begin{figure}[!htb]
\begin{tikzpicture}[x=1pt,y=1pt]
\definecolor{fillColor}{RGB}{255,255,255}
\path[use as bounding box,fill=fillColor,fill opacity=0.00] (0,0) rectangle (252.94,180.67);
\begin{scope}
\path[clip] (  0.00,  0.00) rectangle (252.94,180.67);
\definecolor{drawColor}{RGB}{255,255,255}
\definecolor{fillColor}{RGB}{255,255,255}

\path[draw=drawColor,line width= 0.5pt,line join=round,line cap=round,fill=fillColor] ( -0.00,  0.00) rectangle (252.95,180.68);
\end{scope}
\begin{scope}
\path[clip] ( 39.05, 27.90) rectangle (247.95,175.68);
\definecolor{fillColor}{RGB}{255,255,255}

\path[fill=fillColor] ( 39.05, 27.90) rectangle (247.95,175.68);
\definecolor{drawColor}{gray}{0.92}

\path[draw=drawColor,line width= 0.3pt,line join=round] ( 39.05, 76.34) --
	(247.95, 76.34);

\path[draw=drawColor,line width= 0.3pt,line join=round] ( 39.05,132.61) --
	(247.95,132.61);

\path[draw=drawColor,line width= 0.3pt,line join=round] ( 83.26, 27.90) --
	( 83.26,175.68);

\path[draw=drawColor,line width= 0.3pt,line join=round] (152.70, 27.90) --
	(152.70,175.68);

\path[draw=drawColor,line width= 0.3pt,line join=round] (222.15, 27.90) --
	(222.15,175.68);

\path[draw=drawColor,line width= 0.5pt,line join=round] ( 39.05, 48.21) --
	(247.95, 48.21);

\path[draw=drawColor,line width= 0.5pt,line join=round] ( 39.05,104.48) --
	(247.95,104.48);

\path[draw=drawColor,line width= 0.5pt,line join=round] ( 39.05,160.75) --
	(247.95,160.75);

\path[draw=drawColor,line width= 0.5pt,line join=round] ( 48.54, 27.90) --
	( 48.54,175.68);

\path[draw=drawColor,line width= 0.5pt,line join=round] (117.98, 27.90) --
	(117.98,175.68);

\path[draw=drawColor,line width= 0.5pt,line join=round] (187.42, 27.90) --
	(187.42,175.68);
\definecolor{drawColor}{RGB}{0,0,0}
\definecolor{fillColor}{RGB}{0,0,0}

\path[draw=drawColor,line width= 0.4pt,line join=round,line cap=round,fill=fillColor] ( 48.54,157.71) circle (  1.96);

\path[draw=drawColor,line width= 0.4pt,line join=round,line cap=round,fill=fillColor] ( 69.45,166.43) circle (  1.96);

\path[draw=drawColor,line width= 0.4pt,line join=round,line cap=round,fill=fillColor] ( 81.68,168.96) circle (  1.96);

\path[draw=drawColor,line width= 0.4pt,line join=round,line cap=round,fill=fillColor] ( 90.35,163.35) circle (  1.96);

\path[draw=drawColor,line width= 0.4pt,line join=round,line cap=round,fill=fillColor] ( 97.08,162.32) circle (  1.96);

\path[draw=drawColor,line width= 0.4pt,line join=round,line cap=round,fill=fillColor] (102.58,152.81) circle (  1.96);

\path[draw=drawColor,line width= 0.4pt,line join=round,line cap=round,fill=fillColor] (107.23,146.66) circle (  1.96);

\path[draw=drawColor,line width= 0.4pt,line join=round,line cap=round,fill=fillColor] (111.26,141.09) circle (  1.96);

\path[draw=drawColor,line width= 0.4pt,line join=round,line cap=round,fill=fillColor] (114.81,134.26) circle (  1.96);

\path[draw=drawColor,line width= 0.4pt,line join=round,line cap=round,fill=fillColor] (117.98,124.15) circle (  1.96);

\path[draw=drawColor,line width= 0.4pt,line join=round,line cap=round,fill=fillColor] (120.86,120.06) circle (  1.96);

\path[draw=drawColor,line width= 0.4pt,line join=round,line cap=round,fill=fillColor] (123.48,119.31) circle (  1.96);

\path[draw=drawColor,line width= 0.4pt,line join=round,line cap=round,fill=fillColor] (125.90,111.24) circle (  1.96);

\path[draw=drawColor,line width= 0.4pt,line join=round,line cap=round,fill=fillColor] (128.13,107.83) circle (  1.96);

\path[draw=drawColor,line width= 0.4pt,line join=round,line cap=round,fill=fillColor] (130.21,111.24) circle (  1.96);

\path[draw=drawColor,line width= 0.4pt,line join=round,line cap=round,fill=fillColor] (132.16,103.85) circle (  1.96);

\path[draw=drawColor,line width= 0.4pt,line join=round,line cap=round,fill=fillColor] (133.99, 93.22) circle (  1.96);

\path[draw=drawColor,line width= 0.4pt,line join=round,line cap=round,fill=fillColor] (135.71,100.79) circle (  1.96);

\path[draw=drawColor,line width= 0.4pt,line join=round,line cap=round,fill=fillColor] (137.34, 82.17) circle (  1.96);

\path[draw=drawColor,line width= 0.4pt,line join=round,line cap=round,fill=fillColor] (138.89, 95.34) circle (  1.96);

\path[draw=drawColor,line width= 0.4pt,line join=round,line cap=round,fill=fillColor] (140.36, 88.31) circle (  1.96);

\path[draw=drawColor,line width= 0.4pt,line join=round,line cap=round,fill=fillColor] (141.76, 78.40) circle (  1.96);

\path[draw=drawColor,line width= 0.4pt,line join=round,line cap=round,fill=fillColor] (143.10, 78.40) circle (  1.96);

\path[draw=drawColor,line width= 0.4pt,line join=round,line cap=round,fill=fillColor] (144.39, 88.31) circle (  1.96);

\path[draw=drawColor,line width= 0.4pt,line join=round,line cap=round,fill=fillColor] (145.62, 82.17) circle (  1.96);

\path[draw=drawColor,line width= 0.4pt,line join=round,line cap=round,fill=fillColor] (146.80, 73.95) circle (  1.96);

\path[draw=drawColor,line width= 0.4pt,line join=round,line cap=round,fill=fillColor] (147.94, 73.95) circle (  1.96);

\path[draw=drawColor,line width= 0.4pt,line join=round,line cap=round,fill=fillColor] (149.04, 85.43) circle (  1.96);

\path[draw=drawColor,line width= 0.4pt,line join=round,line cap=round,fill=fillColor] (150.09, 90.89) circle (  1.96);

\path[draw=drawColor,line width= 0.4pt,line join=round,line cap=round,fill=fillColor] (151.12, 78.40) circle (  1.96);

\path[draw=drawColor,line width= 0.4pt,line join=round,line cap=round,fill=fillColor] (152.10, 82.17) circle (  1.96);

\path[draw=drawColor,line width= 0.4pt,line join=round,line cap=round,fill=fillColor] (153.06, 73.95) circle (  1.96);

\path[draw=drawColor,line width= 0.4pt,line join=round,line cap=round,fill=fillColor] (153.99, 68.49) circle (  1.96);

\path[draw=drawColor,line width= 0.4pt,line join=round,line cap=round,fill=fillColor] (154.89, 78.40) circle (  1.96);

\path[draw=drawColor,line width= 0.4pt,line join=round,line cap=round,fill=fillColor] (155.76, 34.61) circle (  1.96);

\path[draw=drawColor,line width= 0.4pt,line join=round,line cap=round,fill=fillColor] (156.61, 61.46) circle (  1.96);

\path[draw=drawColor,line width= 0.4pt,line join=round,line cap=round,fill=fillColor] (157.44, 68.49) circle (  1.96);

\path[draw=drawColor,line width= 0.4pt,line join=round,line cap=round,fill=fillColor] (158.24, 61.46) circle (  1.96);

\path[draw=drawColor,line width= 0.4pt,line join=round,line cap=round,fill=fillColor] (159.03, 34.61) circle (  1.96);

\path[draw=drawColor,line width= 0.4pt,line join=round,line cap=round,fill=fillColor] (159.79, 61.46) circle (  1.96);

\path[draw=drawColor,line width= 0.4pt,line join=round,line cap=round,fill=fillColor] (160.54, 61.46) circle (  1.96);

\path[draw=drawColor,line width= 0.4pt,line join=round,line cap=round,fill=fillColor] (161.26, 68.49) circle (  1.96);

\path[draw=drawColor,line width= 0.4pt,line join=round,line cap=round,fill=fillColor] (161.97, 34.61) circle (  1.96);

\path[draw=drawColor,line width= 0.4pt,line join=round,line cap=round,fill=fillColor] (163.34, 51.55) circle (  1.96);

\path[draw=drawColor,line width= 0.4pt,line join=round,line cap=round,fill=fillColor] (164.01, 34.61) circle (  1.96);

\path[draw=drawColor,line width= 0.4pt,line join=round,line cap=round,fill=fillColor] (164.66, 51.55) circle (  1.96);

\path[draw=drawColor,line width= 0.4pt,line join=round,line cap=round,fill=fillColor] (165.29, 61.46) circle (  1.96);

\path[draw=drawColor,line width= 0.4pt,line join=round,line cap=round,fill=fillColor] (165.91, 34.61) circle (  1.96);

\path[draw=drawColor,line width= 0.4pt,line join=round,line cap=round,fill=fillColor] (166.52, 61.46) circle (  1.96);

\path[draw=drawColor,line width= 0.4pt,line join=round,line cap=round,fill=fillColor] (167.70, 34.61) circle (  1.96);

\path[draw=drawColor,line width= 0.4pt,line join=round,line cap=round,fill=fillColor] (169.94, 34.61) circle (  1.96);

\path[draw=drawColor,line width= 0.4pt,line join=round,line cap=round,fill=fillColor] (171.51, 34.61) circle (  1.96);

\path[draw=drawColor,line width= 0.4pt,line join=round,line cap=round,fill=fillColor] (173.01, 51.55) circle (  1.96);

\path[draw=drawColor,line width= 0.4pt,line join=round,line cap=round,fill=fillColor] (173.49, 34.61) circle (  1.96);

\path[draw=drawColor,line width= 0.4pt,line join=round,line cap=round,fill=fillColor] (174.43, 34.61) circle (  1.96);

\path[draw=drawColor,line width= 0.4pt,line join=round,line cap=round,fill=fillColor] (177.93, 34.61) circle (  1.96);

\path[draw=drawColor,line width= 0.4pt,line join=round,line cap=round,fill=fillColor] (178.75, 34.61) circle (  1.96);

\path[draw=drawColor,line width= 0.4pt,line join=round,line cap=round,fill=fillColor] (184.25, 34.61) circle (  1.96);

\path[draw=drawColor,line width= 0.4pt,line join=round,line cap=round,fill=fillColor] (185.24, 51.55) circle (  1.96);

\path[draw=drawColor,line width= 0.4pt,line join=round,line cap=round,fill=fillColor] (187.42, 34.61) circle (  1.96);

\path[draw=drawColor,line width= 0.4pt,line join=round,line cap=round,fill=fillColor] (191.11, 34.61) circle (  1.96);

\path[draw=drawColor,line width= 0.4pt,line join=round,line cap=round,fill=fillColor] (192.16, 34.61) circle (  1.96);

\path[draw=drawColor,line width= 0.4pt,line join=round,line cap=round,fill=fillColor] (192.67, 34.61) circle (  1.96);

\path[draw=drawColor,line width= 0.4pt,line join=round,line cap=round,fill=fillColor] (193.17, 34.61) circle (  1.96);

\path[draw=drawColor,line width= 0.4pt,line join=round,line cap=round,fill=fillColor] (193.67, 34.61) circle (  1.96);

\path[draw=drawColor,line width= 0.4pt,line join=round,line cap=round,fill=fillColor] (195.10, 34.61) circle (  1.96);

\path[draw=drawColor,line width= 0.4pt,line join=round,line cap=round,fill=fillColor] (195.80, 34.61) circle (  1.96);

\path[draw=drawColor,line width= 0.4pt,line join=round,line cap=round,fill=fillColor] (197.57, 34.61) circle (  1.96);

\path[draw=drawColor,line width= 0.4pt,line join=round,line cap=round,fill=fillColor] (198.63, 34.61) circle (  1.96);

\path[draw=drawColor,line width= 0.4pt,line join=round,line cap=round,fill=fillColor] (219.01, 34.61) circle (  1.96);

\path[draw=drawColor,line width= 0.4pt,line join=round,line cap=round,fill=fillColor] (233.58, 34.61) circle (  1.96);

\path[draw=drawColor,line width= 0.4pt,line join=round,line cap=round,fill=fillColor] (238.45, 34.61) circle (  1.96);
\definecolor{drawColor}{gray}{0.20}

\path[draw=drawColor,line width= 0.5pt,line join=round,line cap=round] ( 39.05, 27.90) rectangle (247.95,175.68);
\end{scope}
\begin{scope}
\path[clip] (  0.00,  0.00) rectangle (252.94,180.67);
\definecolor{drawColor}{gray}{0.30}

\node[text=drawColor,anchor=base east,inner sep=0pt, outer sep=0pt, scale=  0.80] at ( 34.55, 45.45) {0.001};

\node[text=drawColor,anchor=base east,inner sep=0pt, outer sep=0pt, scale=  0.80] at ( 34.55,101.72) {0.010};

\node[text=drawColor,anchor=base east,inner sep=0pt, outer sep=0pt, scale=  0.80] at ( 34.55,158.00) {0.100};
\end{scope}
\begin{scope}
\path[clip] (  0.00,  0.00) rectangle (252.94,180.67);
\definecolor{drawColor}{gray}{0.20}

\path[draw=drawColor,line width= 0.5pt,line join=round] ( 36.55, 48.21) --
	( 39.05, 48.21);

\path[draw=drawColor,line width= 0.5pt,line join=round] ( 36.55,104.48) --
	( 39.05,104.48);

\path[draw=drawColor,line width= 0.5pt,line join=round] ( 36.55,160.75) --
	( 39.05,160.75);
\end{scope}
\begin{scope}
\path[clip] (  0.00,  0.00) rectangle (252.94,180.67);
\definecolor{drawColor}{gray}{0.20}

\path[draw=drawColor,line width= 0.5pt,line join=round] ( 48.54, 25.40) --
	( 48.54, 27.90);

\path[draw=drawColor,line width= 0.5pt,line join=round] (117.98, 25.40) --
	(117.98, 27.90);

\path[draw=drawColor,line width= 0.5pt,line join=round] (187.42, 25.40) --
	(187.42, 27.90);
\end{scope}
\begin{scope}
\path[clip] (  0.00,  0.00) rectangle (252.94,180.67);
\definecolor{drawColor}{gray}{0.30}

\node[text=drawColor,anchor=base,inner sep=0pt, outer sep=0pt, scale=  0.80] at ( 48.54, 17.89) {1};

\node[text=drawColor,anchor=base,inner sep=0pt, outer sep=0pt, scale=  0.80] at (117.98, 17.89) {10};

\node[text=drawColor,anchor=base,inner sep=0pt, outer sep=0pt, scale=  0.80] at (187.42, 17.89) {100};
\end{scope}
\begin{scope}
\path[clip] (  0.00,  0.00) rectangle (252.94,180.67);
\definecolor{drawColor}{RGB}{0,0,0}

\node[text=drawColor,anchor=base,inner sep=0pt, outer sep=0pt, scale=  1.00] at (143.50,  6.94) {Degree};
\end{scope}
\begin{scope}
\path[clip] (  0.00,  0.00) rectangle (252.94,180.67);
\definecolor{drawColor}{RGB}{0,0,0}

\node[text=drawColor,rotate= 90.00,anchor=base,inner sep=0pt, outer sep=0pt, scale=  1.00] at ( 11.89,101.79) {Relative frequency};
\end{scope}
\end{tikzpicture}
         \caption{The same distribution as \cref{fig:log_all_nodes_degree_distribution}, but only including \emph{reachable} (i.\,e. online) nodes.}
        \label{fig:log_online_nodes_degree_distribution}
\end{figure}
Furthermore, there is a significant difference visible especially between China and the U.S.: although most \ac{IPFS} nodes are in China, the vast majority of reachable nodes stems from the U.S.

As stated in \cref{sec:ipfs-nutshell}, \ac{IPFS} supports connections through multiple network layer protocols; \cref{tab:protocols} shows the prevalence of encountered protocols during our crawls.
If a node was reachable through multiple, say IPv4 addresses, we only count it as one occurence of IPv4 to not distort the count.
The majority of nodes support connections through IPv4, followed by IPv6, whereas other protocols are barely used at all.
The protocols ``ipfs'' and ``p2p-circuit'' are connections through \ac{IPFS}' relay nodes, ``dnsaddr'', ``dns4/6'' are DNS-resolvable addresses and ``onion3'' signals TOR capabilities.

\subsection{Overlay Topology}
\label{subsec:graph_properties}

\begin{table}[htb]
  \center
\begin{tabular}{| c | c | c | c | c |}
 \hline
  & Min. & Mean & Median & Max.\\
 \hline
total-degree & 1 & 14.32 & 10.2 & 942.23 \\
 \hline
in-degree & 1 & 7.16 & 3.03 & 935.48 \\
 \hline
out-degree & 0 & 7.16 & 6.48 & 53.56 \\
 \hline
\end{tabular}
   \caption{Average of the degree statistics of all 2400 crawls.}
  \label{tab:degreeStats}
\end{table}

A crawl yields a view on the nodes $V'$ in the buckets $G' = (V', E')$, of non-client nodes, and their connections to each other, \ie $E'$.
As discussed in \cref{sec:overlay_structure}, we obtain a measurement of $V'$ (which is approximately $V$) and $E'$ which is a strict subset of $E$.
Nevertheless, $G' = (V', E')$ is the graph used by \ac{IPFS} to locate data and nodes through the \ac{DHT} and therefore provides a lot of insights. 

Since there is a huge discrepancy between all discovered and the reachable nodes, and since we cannot measure the properties of unreachable nodes, we distinguish these two classes in the analysis.

\cref{fig:log_all_nodes_degree_distribution,fig:log_online_nodes_degree_distribution} depict the log-log in-degree distribution from the first crawl; note that other crawls yielded similar results.
We differentiate between all found nodes (\cref{fig:log_all_nodes_degree_distribution}) and only reachable nodes (\cref{fig:log_online_nodes_degree_distribution}).
The (roughly) straight line in the figure indicates a highly skewed distribution where some peers have very high in-degree (up to 1000) whereas most peers have a fairly small in-degree.
In other words, the in-degree distribution can be approximated by a power-law, hence, the graph can be classified as scale-free, which is in line with prior measurements on other Kademlia systems~\cite{DBLP:conf/iscc/SalahRS14}.

In the following, we specifically focus on the top degree nodes.
Top degree nodes were defined as the \SI{0.05}{\percent} of all nodes, yielding 22.2 nodes on average.
We refrain from defining a number, say the 20 nodes with the highest degree, since this would weigh crawls with fewer nodes differently than crawls with a higher number of total observed nodes.
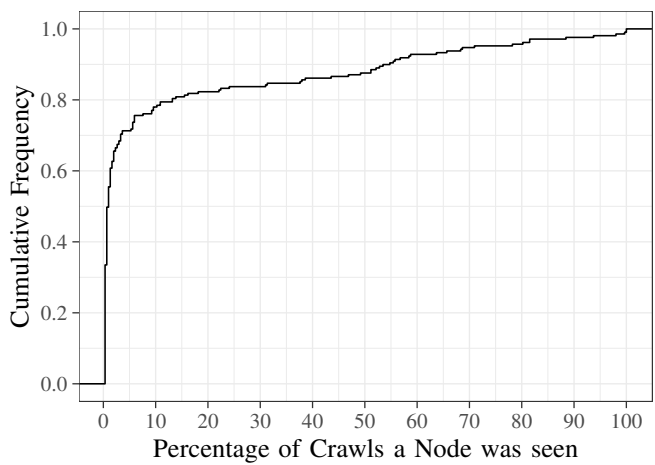
\begin{figure}
\begin{tikzpicture}[x=1pt,y=1pt]
\definecolor{fillColor}{RGB}{255,255,255}
\path[use as bounding box,fill=fillColor,fill opacity=0.00] (0,0) rectangle (252.94,180.67);
\begin{scope}
\path[clip] (  0.00,  0.00) rectangle (252.94,180.67);
\definecolor{drawColor}{RGB}{255,255,255}
\definecolor{fillColor}{RGB}{255,255,255}

\path[draw=drawColor,line width= 0.5pt,line join=round,line cap=round,fill=fillColor] ( -0.00,  0.00) rectangle (252.95,180.68);
\end{scope}
\begin{scope}
\path[clip] ( 31.05, 27.90) rectangle (247.95,175.68);
\definecolor{fillColor}{RGB}{255,255,255}

\path[fill=fillColor] ( 31.05, 27.90) rectangle (247.95,175.68);
\definecolor{drawColor}{gray}{0.92}

\path[draw=drawColor,line width= 0.3pt,line join=round] ( 31.05, 48.05) --
	(247.95, 48.05);

\path[draw=drawColor,line width= 0.3pt,line join=round] ( 31.05, 74.92) --
	(247.95, 74.92);

\path[draw=drawColor,line width= 0.3pt,line join=round] ( 31.05,101.79) --
	(247.95,101.79);

\path[draw=drawColor,line width= 0.3pt,line join=round] ( 31.05,128.65) --
	(247.95,128.65);

\path[draw=drawColor,line width= 0.3pt,line join=round] ( 31.05,155.52) --
	(247.95,155.52);

\path[draw=drawColor,line width= 0.3pt,line join=round] ( 50.15, 27.90) --
	( 50.15,175.68);

\path[draw=drawColor,line width= 0.3pt,line join=round] ( 69.93, 27.90) --
	( 69.93,175.68);

\path[draw=drawColor,line width= 0.3pt,line join=round] ( 89.71, 27.90) --
	( 89.71,175.68);

\path[draw=drawColor,line width= 0.3pt,line join=round] (109.50, 27.90) --
	(109.50,175.68);

\path[draw=drawColor,line width= 0.3pt,line join=round] (129.28, 27.90) --
	(129.28,175.68);

\path[draw=drawColor,line width= 0.3pt,line join=round] (149.06, 27.90) --
	(149.06,175.68);

\path[draw=drawColor,line width= 0.3pt,line join=round] (168.85, 27.90) --
	(168.85,175.68);

\path[draw=drawColor,line width= 0.3pt,line join=round] (188.63, 27.90) --
	(188.63,175.68);

\path[draw=drawColor,line width= 0.3pt,line join=round] (208.41, 27.90) --
	(208.41,175.68);

\path[draw=drawColor,line width= 0.3pt,line join=round] (228.19, 27.90) --
	(228.19,175.68);

\path[draw=drawColor,line width= 0.5pt,line join=round] ( 31.05, 34.61) --
	(247.95, 34.61);

\path[draw=drawColor,line width= 0.5pt,line join=round] ( 31.05, 61.48) --
	(247.95, 61.48);

\path[draw=drawColor,line width= 0.5pt,line join=round] ( 31.05, 88.35) --
	(247.95, 88.35);

\path[draw=drawColor,line width= 0.5pt,line join=round] ( 31.05,115.22) --
	(247.95,115.22);

\path[draw=drawColor,line width= 0.5pt,line join=round] ( 31.05,142.09) --
	(247.95,142.09);

\path[draw=drawColor,line width= 0.5pt,line join=round] ( 31.05,168.96) --
	(247.95,168.96);

\path[draw=drawColor,line width= 0.5pt,line join=round] ( 40.26, 27.90) --
	( 40.26,175.68);

\path[draw=drawColor,line width= 0.5pt,line join=round] ( 60.04, 27.90) --
	( 60.04,175.68);

\path[draw=drawColor,line width= 0.5pt,line join=round] ( 79.82, 27.90) --
	( 79.82,175.68);

\path[draw=drawColor,line width= 0.5pt,line join=round] ( 99.61, 27.90) --
	( 99.61,175.68);

\path[draw=drawColor,line width= 0.5pt,line join=round] (119.39, 27.90) --
	(119.39,175.68);

\path[draw=drawColor,line width= 0.5pt,line join=round] (139.17, 27.90) --
	(139.17,175.68);

\path[draw=drawColor,line width= 0.5pt,line join=round] (158.95, 27.90) --
	(158.95,175.68);

\path[draw=drawColor,line width= 0.5pt,line join=round] (178.74, 27.90) --
	(178.74,175.68);

\path[draw=drawColor,line width= 0.5pt,line join=round] (198.52, 27.90) --
	(198.52,175.68);

\path[draw=drawColor,line width= 0.5pt,line join=round] (218.30, 27.90) --
	(218.30,175.68);

\path[draw=drawColor,line width= 0.5pt,line join=round] (238.09, 27.90) --
	(238.09,175.68);
\definecolor{drawColor}{RGB}{0,0,0}

\path[draw=drawColor,line width= 0.6pt,line join=round] ( 31.05, 34.61) --
	( 40.91, 34.61) --
	( 40.91, 79.61) --
	( 41.56, 79.61) --
	( 41.56,101.46) --
	( 42.22,101.46) --
	( 42.22,109.18) --
	( 42.87,109.18) --
	( 42.87,116.25) --
	( 43.52,116.25) --
	( 43.52,118.82) --
	( 44.17,118.82) --
	( 44.17,122.68) --
	( 44.83,122.68) --
	( 44.83,123.96) --
	( 45.48,123.96) --
	( 45.48,125.25) --
	( 46.13,125.25) --
	( 46.13,126.53) --
	( 46.79,126.53) --
	( 46.79,129.10) --
	( 47.44,129.10) --
	( 47.44,130.39) --
	( 50.70,130.39) --
	( 50.70,131.03) --
	( 51.36,131.03) --
	( 51.36,133.60) --
	( 52.01,133.60) --
	( 52.01,136.18) --
	( 55.27,136.18) --
	( 55.27,136.82) --
	( 58.54,136.82) --
	( 58.54,138.10) --
	( 59.19,138.10) --
	( 59.19,139.39) --
	( 60.50,139.39) --
	( 60.50,140.03) --
	( 61.80,140.03) --
	( 61.80,141.32) --
	( 66.37,141.32) --
	( 66.37,142.60) --
	( 67.68,142.60) --
	( 67.68,143.25) --
	( 70.94,143.25) --
	( 70.94,143.89) --
	( 72.25,143.89) --
	( 72.25,144.53) --
	( 76.17,144.53) --
	( 76.17,145.17) --
	( 84.00,145.17) --
	( 84.00,145.82) --
	( 84.65,145.82) --
	( 84.65,146.46) --
	( 87.92,146.46) --
	( 87.92,147.10) --
	(101.63,147.10) --
	(101.63,147.75) --
	(102.28,147.75) --
	(102.28,148.39) --
	(114.69,148.39) --
	(114.69,149.03) --
	(115.34,149.03) --
	(115.34,149.67) --
	(116.65,149.67) --
	(116.65,150.32) --
	(126.44,150.32) --
	(126.44,150.96) --
	(132.97,150.96) --
	(132.97,151.60) --
	(137.54,151.60) --
	(137.54,152.25) --
	(141.46,152.25) --
	(141.46,153.53) --
	(143.42,153.53) --
	(143.42,154.17) --
	(144.72,154.17) --
	(144.72,154.82) --
	(146.03,154.82) --
	(146.03,155.46) --
	(148.64,155.46) --
	(148.64,156.10) --
	(149.94,156.10) --
	(149.94,156.74) --
	(150.60,156.74) --
	(150.60,157.39) --
	(152.56,157.39) --
	(152.56,158.03) --
	(155.82,158.03) --
	(155.82,158.67) --
	(156.47,158.67) --
	(156.47,159.32) --
	(166.27,159.32) --
	(166.27,159.96) --
	(170.18,159.96) --
	(170.18,160.60) --
	(175.41,160.60) --
	(175.41,161.24) --
	(176.06,161.24) --
	(176.06,161.89) --
	(180.63,161.89) --
	(180.63,162.53) --
	(194.99,162.53) --
	(194.99,163.17) --
	(198.91,163.17) --
	(198.91,163.82) --
	(201.52,163.82) --
	(201.52,165.10) --
	(215.23,165.10) --
	(215.23,165.74) --
	(225.68,165.74) --
	(225.68,166.39) --
	(234.17,166.39) --
	(234.17,167.03) --
	(237.43,167.03) --
	(237.43,167.67) --
	(238.09,167.67) --
	(238.09,168.96) --
	(247.95,168.96) --
	(247.95,168.96);
\definecolor{drawColor}{gray}{0.20}

\path[draw=drawColor,line width= 0.5pt,line join=round,line cap=round] ( 31.05, 27.90) rectangle (247.95,175.68);
\end{scope}
\begin{scope}
\path[clip] (  0.00,  0.00) rectangle (252.94,180.67);
\definecolor{drawColor}{gray}{0.30}

\node[text=drawColor,anchor=base east,inner sep=0pt, outer sep=0pt, scale=  0.80] at ( 26.55, 31.86) {0.0};

\node[text=drawColor,anchor=base east,inner sep=0pt, outer sep=0pt, scale=  0.80] at ( 26.55, 58.73) {0.2};

\node[text=drawColor,anchor=base east,inner sep=0pt, outer sep=0pt, scale=  0.80] at ( 26.55, 85.60) {0.4};

\node[text=drawColor,anchor=base east,inner sep=0pt, outer sep=0pt, scale=  0.80] at ( 26.55,112.47) {0.6};

\node[text=drawColor,anchor=base east,inner sep=0pt, outer sep=0pt, scale=  0.80] at ( 26.55,139.33) {0.8};

\node[text=drawColor,anchor=base east,inner sep=0pt, outer sep=0pt, scale=  0.80] at ( 26.55,166.20) {1.0};
\end{scope}
\begin{scope}
\path[clip] (  0.00,  0.00) rectangle (252.94,180.67);
\definecolor{drawColor}{gray}{0.20}

\path[draw=drawColor,line width= 0.5pt,line join=round] ( 28.55, 34.61) --
	( 31.05, 34.61);

\path[draw=drawColor,line width= 0.5pt,line join=round] ( 28.55, 61.48) --
	( 31.05, 61.48);

\path[draw=drawColor,line width= 0.5pt,line join=round] ( 28.55, 88.35) --
	( 31.05, 88.35);

\path[draw=drawColor,line width= 0.5pt,line join=round] ( 28.55,115.22) --
	( 31.05,115.22);

\path[draw=drawColor,line width= 0.5pt,line join=round] ( 28.55,142.09) --
	( 31.05,142.09);

\path[draw=drawColor,line width= 0.5pt,line join=round] ( 28.55,168.96) --
	( 31.05,168.96);
\end{scope}
\begin{scope}
\path[clip] (  0.00,  0.00) rectangle (252.94,180.67);
\definecolor{drawColor}{gray}{0.20}

\path[draw=drawColor,line width= 0.5pt,line join=round] ( 40.26, 25.40) --
	( 40.26, 27.90);

\path[draw=drawColor,line width= 0.5pt,line join=round] ( 60.04, 25.40) --
	( 60.04, 27.90);

\path[draw=drawColor,line width= 0.5pt,line join=round] ( 79.82, 25.40) --
	( 79.82, 27.90);

\path[draw=drawColor,line width= 0.5pt,line join=round] ( 99.61, 25.40) --
	( 99.61, 27.90);

\path[draw=drawColor,line width= 0.5pt,line join=round] (119.39, 25.40) --
	(119.39, 27.90);

\path[draw=drawColor,line width= 0.5pt,line join=round] (139.17, 25.40) --
	(139.17, 27.90);

\path[draw=drawColor,line width= 0.5pt,line join=round] (158.95, 25.40) --
	(158.95, 27.90);

\path[draw=drawColor,line width= 0.5pt,line join=round] (178.74, 25.40) --
	(178.74, 27.90);

\path[draw=drawColor,line width= 0.5pt,line join=round] (198.52, 25.40) --
	(198.52, 27.90);

\path[draw=drawColor,line width= 0.5pt,line join=round] (218.30, 25.40) --
	(218.30, 27.90);

\path[draw=drawColor,line width= 0.5pt,line join=round] (238.09, 25.40) --
	(238.09, 27.90);
\end{scope}
\begin{scope}
\path[clip] (  0.00,  0.00) rectangle (252.94,180.67);
\definecolor{drawColor}{gray}{0.30}

\node[text=drawColor,anchor=base,inner sep=0pt, outer sep=0pt, scale=  0.80] at ( 40.26, 17.89) {0};

\node[text=drawColor,anchor=base,inner sep=0pt, outer sep=0pt, scale=  0.80] at ( 60.04, 17.89) {10};

\node[text=drawColor,anchor=base,inner sep=0pt, outer sep=0pt, scale=  0.80] at ( 79.82, 17.89) {20};

\node[text=drawColor,anchor=base,inner sep=0pt, outer sep=0pt, scale=  0.80] at ( 99.61, 17.89) {30};

\node[text=drawColor,anchor=base,inner sep=0pt, outer sep=0pt, scale=  0.80] at (119.39, 17.89) {40};

\node[text=drawColor,anchor=base,inner sep=0pt, outer sep=0pt, scale=  0.80] at (139.17, 17.89) {50};

\node[text=drawColor,anchor=base,inner sep=0pt, outer sep=0pt, scale=  0.80] at (158.95, 17.89) {60};

\node[text=drawColor,anchor=base,inner sep=0pt, outer sep=0pt, scale=  0.80] at (178.74, 17.89) {70};

\node[text=drawColor,anchor=base,inner sep=0pt, outer sep=0pt, scale=  0.80] at (198.52, 17.89) {80};

\node[text=drawColor,anchor=base,inner sep=0pt, outer sep=0pt, scale=  0.80] at (218.30, 17.89) {90};

\node[text=drawColor,anchor=base,inner sep=0pt, outer sep=0pt, scale=  0.80] at (238.09, 17.89) {100};
\end{scope}
\begin{scope}
\path[clip] (  0.00,  0.00) rectangle (252.94,180.67);
\definecolor{drawColor}{RGB}{0,0,0}

\node[text=drawColor,anchor=base,inner sep=0pt, outer sep=0pt, scale=  1.00] at (139.50,  6.94) {Percentage of Crawls a Node was seen};
\end{scope}
\begin{scope}
\path[clip] (  0.00,  0.00) rectangle (252.94,180.67);
\definecolor{drawColor}{RGB}{0,0,0}

\node[text=drawColor,rotate= 90.00,anchor=base,inner sep=0pt, outer sep=0pt, scale=  1.00] at ( 11.89,101.79) {Cumulative Frequency};
\end{scope}
\end{tikzpicture}
   \caption{ECDF of how often the same nodes were within the top-degree nodes.}
  \label{fig:maxdegecdf}
\end{figure}
\cref{fig:maxdegecdf} depicts a cumulative distribution of how many times the nodes with the highest degree were seen in the crawls.
If a node was in the set of highest-degree nodes in one run but not in other runs, its ``percentage seen'' would be $\frac{1}{\text{\# of crawls}} = \frac{1}{2400} = 0.0033$ or \SI{0.33}{\percent}.
On the other extreme, if a node was within the highest degree nodes in every crawl, its percentage seen would be a \SI{100}{\percent}.

The cumulative distribution in \cref{fig:maxdegecdf} shows a high churn within the highest degree nodes: approximately \SI{80}{\percent} of nodes were only present in \SI{10}{\percent} of the crawls.
Only a few nodes have a high degree in the majority of all crawls; these nodes are the bootstrap nodes along a handful of others.

\section{Conclusion and Future Work}
\label{sec:conclusion}

Founded in an analysis of network-layer mechanisms actually in use in the current \ac{IPFS} implementation,
we've presented several approaches towards mapping the \ac{IPFS} overlay network.
Using monitoring nodes and a \SI{6.98}{\day} crawl of the \ac{IPFS} \ac{DHT},
we obtain a holistic view on \ac{IPFS}' node population as well as indicators about the larger overlay structure.
Of the 44474 nodes that we found during each crawl on average, only \SI{6.55}{\percent} responded to our connection attempts.
This, among other indicators, hints at the fact that a majority of nodes is operated by private individuals.
Lastly, we found that due to the combination of \ac{DHT}-based and broadcast-based content retrieval, \ac{IPFS} is a hybrid between structured and unstructured overlays; sacrificing performance for improving robustness.

Towards future works,
we observe that our monitoring nodes can be extended to also record data request patterns (similarly to studies conducted in the Bitcoin peer-to-peer network~\cite{neudecker2016timing}).
As \ac{IPFS} nodes broadcast each sought data item to \emph{each} of their connected peers (cf. \cref{subsec:bitswap_exchange}),
a node that is connected to every other node could monitor nearly \emph{all} data requests in the network.
While enabling insights into the "filesystem" side of \ac{IPFS},
the success of such a measurement approach
would also result in dire implications for the query privacy of \ac{IPFS} users.

\printbibliography
\end{document}